\documentclass[12pt]{article}

\usepackage{amsmath}
\usepackage{amssymb}
\usepackage{amsthm}
\usepackage{graphicx}
\usepackage{ifthen}
\usepackage{verbatim}
\usepackage{color}
\usepackage{leftidx}
\usepackage{bm}
\usepackage{titlefoot}
\usepackage{authblk}

\theoremstyle{plain}
\newtheorem{theorem}{Theorem}
\newtheorem{proposition}{Proposition}[section]

\theoremstyle{definition}

\theoremstyle{remark}
\newtheorem{remark}{Remark}[section]

\newcommand{\bst}{\boldsymbol{t}}
\newcommand{\calL}{\mathcal{L}}
\newcommand{\calT}{\mathcal{T}}

\numberwithin{equation}{section}

\allowdisplaybreaks 

\makeatletter
  \def\thefootnote{\ifnum\c@footnote>\z@\leavevmode\lower.5ex%
      \hbox{$^{\@arabic\c@footnote)}$}\fi}
\makeatother

\begin{document}

\title{%
\textbf{\Large 
Three-partition Hodge integrals \\[-0mm]  
         and \\[-0mm]
the topological vertex \\[5mm]}
}
        
\author{%
{\large 
Toshio Nakatsu\footnote{E-mail: \texttt{nakatsu@mpg.setsunan.ac.jp}} } \\ 
\vspace{-4mm}
\textit{\normalsize Institute for Fundamental Sciences, Setsunan University} \\
\vspace{-1mm}
\textit{\normalsize 17-8 Ikeda Nakamachi, Neyagawa, Osaka 572-8508, Japan}
\and 
{\large 
Kanehisa Takasaki\footnote{E-mail: \texttt{takasaki@math.kindai.ac.jp}}} \\
\vspace{-4mm}
\textit{\normalsize Department of Mathematics, Kindai University} \\
\vspace{-1mm}
\textit{\normalsize 3-4-1 Kowakae, Higashi-Osaka, Osaka 577-8502, Japan}
}
 
\date{} 
 
\maketitle

\begin{abstract}
A conjecture on the relation between the cubic Hodge integrals 
and the topological vertex in topological string theory is resolved. 
A central role is played by the notion 
of generalized shift symmetries in a fermionic realization 
of the two-dimensional quantum torus algebra.  
These algebraic relations of operators in the fermionic Fock space 
are used to convert generating functions of the cubic Hodge integrals 
and the topological vertex to each other.  
As a byproduct, 
the generating function of the cubic Hodge integrals 
at special values of the parameters $\overrightarrow{w}$ 
therein is shown to be a tau function 
of the generalized KdV (aka Gelfand-Dickey) hierarchies.  
\end{abstract}

\date{December, 2018} 

\newpage

\section{Introduction}

Let $\mathcal{M}_{g,n}$ denote the moduli space of 
connected complex algebraic curves of genus $g$ with $n$ marked points 
and $\overline{\mathcal{M}}_{g,n}$ the Deligne-Mumford compactification. 
Marked points on an algebraic curve $C$ of genus $g$ are referred to 
as $z_1,\dots,z_n$.  
The Hodge integrals are integrals on $\overline{\mathcal{M}}_{g,n}$ 
of the form 
$$ 
 \int_{\overline{\mathcal{M}}_{g, n}}
 \psi_1^{i_1}\cdots\psi_n^{i_n}\,\lambda_1^{j_1}\cdots\lambda_g^{j_g}, 
$$
where $\psi_i =c_1(\mathbb{L}_i)$ is the first Chern class 
of the bundle $\mathbb{L}_i$ of the cotangent space of $C$  
at the $i$-th marked point $z_i$, and $\lambda_j =c_j(\mathbb{E})$ 
is the $j$-th Chern class of  the Hodge bundle $\mathbb{E}$ 
over $\overline{\mathcal{M}}_{g,n}$.  
These Hodge integrals are building blocks 
of the localization computation of Gromov-Witten invariants 
\cite{Graber_Pandharipande_1999}. 
The localization formula expresses those invariants 
as a sum of weighted graphs \cite{Kontsevich_1995}.  
Integrals of the form 
\begin{align*}
    \int_{\overline{\mathcal{M}}_{g,n}} 
    \frac{\prod_{i=1}^s \bigwedge^\vee_g(u_i)}
           {\prod_{j=1}^{n}\bigl(1-z_j \psi_j\bigr)},
\end{align*}
where $\mbox{$\bigwedge^\vee_g$}(u)$ is the characteristic polynomial 
\begin{align*}
   \mbox{$\bigwedge^\vee_g$}(u) 
      = \sum_{r=0}^g(-1)^r\lambda_ru^{g-r} 
      = u^g-\lambda_1u^{g-1}+\,\cdots\,+(-1)^g\lambda_g, 
\end{align*}
arise as the vertex weights 
in the localization computation of Gromov-Witten invariants 
of an $s$-dimensional target space. 

Li, Liu, Liu and Zhou \cite{Li_Liu_Liu_Zhou_2009} 
reformulated the topological vertex in topological string theory \cite{AKMV} 
on the basis of formal relative Gromov-Witten theory 
of non-compact Calabi-Yau threefolds.  
The topological vertex is a diagrammatic method 
to compute all-genus A-model topological string amplitudes 
on smooth non-compact toric Calabi-Yau threefolds.  
It is somewhat confusing that the vertex weight itself is called the topological vertex. 
All building blocks of the vertex weight are genuinely combinatorial 
objects such as partitions, the Littlewood-Richardson numbers 
and the Schur functions.  

The geometric approach of Li et al. stems from 
their proof of the Mari\~{n}o-Vafa conjecture and its generalization 
\cite{Liu_Liu_Zhou_2004, Liu_Liu_Zhou_2003, Liu_Liu_Zhou_2007}.
\footnote{
The Mari\~{n}o-Vafa conjecture was also resolved 
by Okounkov and Pandharipande using a quite different method.
} 
The Mari\~{n}o-Vafa conjecture \cite{Marino_Vafa_2001}  
is a variation of the so called ELSV formula \cite{ELSV_2001}. 
The ELSV formula expresses the Hurwitz numbers 
of ramified coverings of $\mathbb{P}^1$ in terms of linear ($s=1$) Hodge integrals.  
The Mari\~{n}o-Vafa conjecture claims a similar relation 
between certain open Gromov-Witten invariants and 
cubic ($s=3$) Hodge integrals.  Actually, 
the open Gromov-Witten invariants considered therein 
are related to a special case of the topological vertex.  

The approach to the full (i.e., three-partition) topological vertex, 
just like the proof of the Mari\~{n}o-Vafa conjecture, 
starts from a generating function of cubic Hodge integrals.  
The cubic Hodge integrals are labelled by a triple 
$\overrightarrow{\mu} = (\mu^{(1)},\mu^{(2)},\mu^{(3)})$ 
of partitions and depend on a triple 
$\overrightarrow{w} = (w_1,w_2,w_3)$ of parameters.  
Li et al. employ various skillful ideas to derive 
a combinatorial expression of the topological vertex 
from the generating function.  This expression, however, 
takes a form apparently different from the topological vertex 
of Aganagic et al.  Showing several pieces of evidence, 
Li et al. conjecture that the two expressions 
of the topological vertex are equivalent.  

In this paper, 
we prove the equivalence of the two different expressions of the topological vertex.  
Our strategy is to construct a generating function 
of the topological vertex of Aganagic et al. 
(modified to depend on the parameters $\overrightarrow{w}$ of the cubic Hodge integrals), 
and to show that it coincides with (an exponentiated version of) 
the generating function of Li et al.  
To this end, we make use of tools developed 
in a series of our work on the melting crystal models 
\cite{Nakatsu_Takasaki_CMP_2009,Nakatsu_Takasaki_ASPS}
and topological string theory 
\cite{Takasaki_Nakatsu_JPA_2016,Takasaki_2014,Takasaki_Nakatsu_SIGMA_2017}. 
A central role is played by the notion 
of generalized shift symmetries \cite{Nakatsu_Takasaki_2019} 
in a fermionic realization of the two-dimensional quantum torus algebra.  
We start from a fermionic representation of the generating function 
of the topological vertex.  The generalized shift symmetries enable us 
to convert this fermionic representation to the generating function 
of cubic Hodge integrals of Li et al.  This implies the equivalence 
of the two expressions of the topological vertex. 

As a byproduct, 
we find a new link between the topological vertex and integrable hierarchies.  
Just after the first proposal \cite{AKMV}, 
particular generating functions of the topological vertex 
were pointed out to be tau functions of integrable hierarchies 
such as the KP and 2D Toda hierarchies \cite{ADKMV,Zhou_2010}. 
We show that the generating function of the cubic Hodge integrals 
at special values of the parameters $\overrightarrow{w}$ is a tau function 
of the generalized KdV (aka Gelfand-Dickey) hierarchies.  
Such a link with the generalized KdV hierarchies  
seems to be unknown in the past literature. 

Let us explain these backgrounds and results in more detail. 

\subsection{Partitions, representations of symmetric group and the Schur functions} 

Our notations for partitions, representations of symmetric groups and 
the Schur functions are mostly borrowed from Macdonald's book \cite{Macdonald_book}. 
Let us recall these notions and basic facts.

A partition is a sequence of non-negative integers 
$$ 
   \mu =(\mu_i)_{i=1}^\infty =(\mu_1,\mu_2,\,\dots\,) 
$$  
satisfying $\mu_i\geq\mu_{i +1}$ for all $i\geq 1$. 
The number of the non-zero $\mu_i$ is the length of $\mu$, denoted by $l(\mu)$. 
The sum of the non-zero $\mu_i$ is the weight of $\mu$, 
denoted by $|\mu|=\mu_1+\mu_2+\cdots$. 
Partitions have another expression which indicates the number of times 
each non-negative integers occurs in a partition
$$ 
   \mu =(1^{m_1}2^{m_2}\dots\,), 
$$ 
where $m_i=m_i(\mu)$ denotes the number of $i$ occurs in $\mu$.  
The automorphism group of $\mu$, denoted by $\mbox{Aut}(\mu)$, 
consists of possible permutations among the non-zero $\mu_i$'s that leave $\mu$. 
The number of elements in $\mbox{Aut}(\mu)$ is 
$$ 
    |\mbox{Aut}(\mu)| =\prod_{i=1}^\infty m_i(\mu)!. 
$$ 
Note that partitions are identified with the Young diagrams.  
The size of the Young diagram $\mu$ is $|\mu|$, 
which is the total number of boxes of the diagram, 
and $l(\mu)$ is the height of the diagram. 
The conjugate (or transpose) of $\mu$ is the partition $\ltrans{\,\mu}$ 
whose diagram is the transpose of the diagram $\mu$.

If $|\mu|=d$, we say that $\mu$ is a partition of $d$. 
Each partition of $d$ corresponds to a conjugacy class 
of the $d$-th symmetric group $S_d$. 
Let $C(\mu)$ denote the conjugacy class determined by a partition $\mu$ of $d$. 
It is determined by the cycle type $\mu$ of a representative $\sigma \in S_d$ 
of $C(\mu)$ as 
$$ 
  \sigma =(1,\dots,\mu_1)(\mu_1+1,\dots,\mu_1+\mu_2) 
               \cdots (\mu_1+\dots+\mu_{l-1},\dots,d), 
$$
where $l=l(\mu)$ and $(j_1,\dots, j_m)$ means the cyclic permutation 
sending $j_1\to j_2\to\cdots\to j_m\to j_1$. 
The number of elements in $C(\mu)$ is 
$$
   |C(\mu)| =\frac{d!}{z_\mu}, \quad z_\mu 
                 =\prod_{i=1}^\infty m_i(\mu)!\,i^{m_i(\mu)}. 
$$
Each partition $\mu$ of $d$ determines 
the irreducible representation $(\rho_\mu,V_\mu)$ of $S_d$. 
$\chi_\mu$ denotes the character $\mbox{Tr}_{V_\mu}\rho_\mu$. 
The value of $\chi_\mu$ on the conjugacy class $C(\nu)\subset S_d$, 
denoted by $\chi_\mu (\nu)=\chi_\mu\left(C(\nu)\right)$, 
can be computed by the Frobenius formula 
\begin{eqnarray}
    s_\mu (\bm{x}) =\sum_{|\nu|=d}\frac{\chi_{\mu}(\nu)}{z_\nu}p_\nu, \quad
    \bm{x} =(x_i)_{i=1}^\infty =(x_1,x_2,\dots\,), 
    \label{Frobenius}
\end{eqnarray}
where $p_\nu$'s are the monomials 
$$ 
   p_\nu =p_{\nu_1}p_{\nu_2}\dots 
$$ 
of the power sums 
$$
   p_k =\sum_{i=1}^\infty x_i^k,\quad k=1,2,\dots\,,  
$$
and $s_\mu(\bm{x})$ denotes the Schur function. 
(\ref{Frobenius}) has the inversion formula  
\begin{eqnarray}
     p_\mu =\sum_{|\nu|=d}\chi_\nu(\mu)s_\nu(\bm{x}).  
     \label{inverse_Frobenius}
\end{eqnarray}

Let us recall the Jacobi-Trudi formula 
\begin{eqnarray}
     s_\mu(\bm{x}) 
             =\mbox{det}\left(h_{\mu_i-i+j}(\bm{x})\right)_{i, j=1}^n,
     \label{JT_formula_Schur}
\end{eqnarray}
where the length of $\mu=(\mu_i)_{i=1}^\infty$ is not larger than $n$
 and $h_m(\bm{x})$'s are the complete symmetric functions 
 defined by the generating function 
$$
   \sum_{m=0}^\infty h_m(\bm{x})z^m =\prod_{i=1}^\infty (1-x_iz)^{-1}. 
$$
Likewise, 
the skew Schur functions have a determinant formula 
similar to (\ref{JT_formula_Schur}). 
Let $\nu =(\nu_i)_{i=1}^\infty$ be a partition such that 
$\mu\supset\nu$, {\it \small  i.e.}, $\mu_i \geq \nu_i$ for all $i$. 
The determinant formula reads   
\begin{eqnarray}
    s_{\mu /\nu}(\bm{x}) 
         =\mbox{det}\left(h_{\mu_i-\nu_j-i+j }(\bm{x})\right)_{i, j=1}^n, 
    \label{JT_formula_skewSchur}
\end{eqnarray} 
where $\mu /\nu$ represents the skew diagram obtained by 
removing the Young diagram $\nu$ from the larger one $\mu$.  

The Littlewood-Richardson numbers $c_{\mu\nu}^{\eta}$ are 
non-negative integers which are determined by the relation 
\begin{eqnarray}
    s_\mu(\bm{x})s_\nu(\bm{x}) 
               =\sum_{\eta\in\mathcal{P}}c_{\mu\nu}^{\eta}s_\eta(\bm{x}), 
    \label{def_LR}
\end{eqnarray}
where the sum with respect to $\eta$ ranges over 
the set $\mathcal{P}$ of all partitions, and is a finite sum 
because $c_{\mu\nu}^{\eta} =0$ unless 
$|\eta| =|\mu|+|\nu|$, $\eta\supset\mu$ and $\eta\supset\nu$.
The skew Schur functions can be expressed in a linear combination 
of the Schur functions weighted by the Littlewood-Richardson numbers as  
\begin{eqnarray}
     s_{\mu/\nu}(\bm{x}) 
     =\sum_{\eta\in\mathcal{P}}c_{\nu\eta}^{\mu}s_\eta(\bm{x}).
    \label{skewSchur_LR}
\end{eqnarray} 

\subsection{Three-partition Hodge integrals}

The three-partition Hodge integrals are Hodge integrals that depend on 
a triple of partitions in a specific manner.  
For a triple of partitions 
$\overrightarrow{\mu} =(\mu^{(1)},\mu^{(2)},\mu^{(3)})$ $\in$ 
$\mathcal{P}^3=\mathcal{P}\times\mathcal{P}\times\mathcal{P}$, 
we use the notations  
\begin{align*}
    l(\overrightarrow{\mu}) =\sum_{a=1}^3l(\mu^{(a)}),\quad
    \mbox{Aut}(\overrightarrow{\mu}) =\prod_{a=1}^3\mbox{Aut}(\mu^{(a)}).
\end{align*}
$l(\overrightarrow{\mu})$ marked points on an algebraic curve of genus $g$ 
are referred to as $z^{(1)}_1,\dots,z^{(1)}_{l(\mu^{(1)})}$, 
$z^{(2)}_1,\dots,z^{(2)}_{l(\mu^{(2)})}$ and $z^{(3)}_1,\dots,z^{(3)}_{l(\mu^{(3)})}$. 
Cotangent lines of curves at the marked point $z^{(a)}_i$ are glued together 
and form the complex line bundle $\mathbb{L}^{(a)}_i$ 
over $\overline{\mathcal{M}}_{g,l(\overrightarrow{\mu})}$.   

Let $\overrightarrow{w} =(w_1,w_2,w_3)$ be a triple of variables 
which satisfy the condition $w_1+w_2+w_3=0$.  
The indices of those variables are understood to be cyclic as $w_{i+3}\equiv w_i$.  
For $\overrightarrow{\mu}\in\mathcal{P}_+^3 
=\mathcal{P}^3-\{(\emptyset,\emptyset,\emptyset)\}$, 
the three-partition Hodge integrals are defined by 
\begin{equation}  
\begin{aligned}
    G_{g,\overrightarrow{\mu}}(\overrightarrow{w}) 
    & = \frac{(\sqrt{-1})^{l(\overrightarrow{\mu})}}
                  {|\mbox{Aut}(\overrightarrow{\mu})|}
          \prod_{a=1}^3 \prod_{i=1}^{l(\mu^{(a)})} 
          \biggl\{ 
               w_{a+1} \prod_{j=1}^{\mu^{(a)}_i} 
               \Bigl(1+\frac{\mu_i^{(a)}w_{a+1}}{jw_a}\Bigr) 
          \biggr\} 
    \\[1.5mm]
    & \quad\,\times\, 
    \int_{\overline{\mathcal{M}}_{g,l(\overrightarrow{\mu})}} 
    \prod_{a=1}^3 
    \frac{\bigwedge^\vee_g(w_a)w_a^{l(\mu^{(a)})-1}}
           {\prod_{i=1}^{l(\mu^{(a)})}\bigl(w_a-\mu_i^{(a)}\psi_i^{(a)}\bigr)},
\end{aligned}
\label{def_3-partition Hodge integral} 
\end{equation}
where $\psi^{(a)}_i =c_1(\mathbb{L}^{(a)}_i)$ are 
the $\psi$-classes associated with the marked points $z_i^{(a)}$. 

We introduce generating functions of these three-partition Hodge integrals. 
Let $p^{(1)} =(p^{(1)}_k)_{k=1}^\infty$, $p^{(2)} =(p^{(2)}_k)_{k=1}^\infty$ 
and $p^{(3)} =(p^{(3)}_k)_{k=1}^\infty$ be formal variables, 
denoted collectively by $\overrightarrow{p} =(p^{(1)},p^{(2)},p^{(3)})$. 
We put $p^{(a)}_\mu =\prod_{i=1}^{l(\mu)} p^{(a)}_{\mu_i}$ 
for a non-zero partition $\mu=(\mu_i)_{i=1}^\infty$, 
and $p^{(a)}_\emptyset =1$ for the zero partition $\emptyset$. 
The generating functions are defined by 
\begin{align}
   G_{\overrightarrow{\mu}}\left(\lambda;\overrightarrow{w}\right)
   & = \sum_{g=0}^\infty\lambda^{2g-2+l(\overrightarrow{\mu})}
                   G_{g,\overrightarrow{\mu}}\left(\overrightarrow{w}\right),  
 \label{connected_G_genus} \\[1mm]
   G\left(\lambda;\overrightarrow{p};\overrightarrow{w}\right)
   & = \sum_{\overrightarrow{\mu}\in\mathcal{P}^3_+}
                   G_{\overrightarrow{\mu}}\left(\lambda;\overrightarrow{w}\right)
                           \prod_{a=1}^3p^{(a)}_{\mu^{(a)}}, 
 \label{connected_G} 
\end{align}
where $\lambda$ amounts to the string coupling constant.

By degree counting of the RHS of (\ref{def_3-partition Hodge integral}),  
three-partition Hodge integrals turn out to be homogeneous of degree 0 
with respect to $\overrightarrow{w}$: 
\begin{align*}
   G_{g,\overrightarrow{\mu}}\left(t\overrightarrow{w}\right) 
        = G_{g,\overrightarrow{\mu}}\left(\overrightarrow{w}\right), \quad 
               t\overrightarrow{w}=(tw_1,tw_2,tw_3),\quad 
               t\neq 0.
\end{align*}
This implies that 
\begin{align*}
   G_{\overrightarrow{\mu}}\left(\lambda;t\overrightarrow{w}\right)
        = G_{\overrightarrow{\mu}}\left(\lambda;\overrightarrow{w}\right), \quad
   G\left(\lambda;\overrightarrow{p};t\overrightarrow{w}\right)
        = G\left(\lambda;\overrightarrow{p};\overrightarrow{w}\right).
\end{align*}
Therefore, $\overrightarrow{w}$ can be chosen as  
\begin{eqnarray}
     \overrightarrow{w} = (1,\,\tau,\,-1-\tau) 
  \label{tau}
\end{eqnarray}
where $\tau$ is a new variable. 
For $\overrightarrow{w}$ of the form (\ref{tau}), 
the three-partition Hodge integrals are expressed in a shortened form as   
$G_{g,\overrightarrow{\mu}}(\tau) = G_{g,\overrightarrow{\mu}}(1,\tau,-1-\tau)$. 
The same abbreviation is also used for the generating functions 
(\ref{connected_G_genus}) and (\ref{connected_G}) as 
\begin{align*}
   G_{\overrightarrow{\mu}}\left(\lambda;\tau\right)
        = G_{\overrightarrow{\mu}}\left(\lambda;1,\tau ,-1-\tau\right),\quad
   G\left(\lambda;\overrightarrow{p};\tau\right)
       = G\left(\lambda;\overrightarrow{p};1,\tau ,-1-\tau\right). 
\end{align*}

The generating function (\ref{connected_G}) can be expanded 
in an infinite series of the Schur functions 
by the inversion formula (\ref{inverse_Frobenius}). 
The disconnected version of (\ref{connected_G}) turns out to 
take the specific form \cite{Li_Liu_Liu_Zhou_2009} 
\begin{eqnarray}
     \exp\left( G(\lambda;\overrightarrow{p};\overrightarrow{w}) \right) 
       = \sum_{\overrightarrow{\mu}\in\mathcal{P}^3} 
                  \widetilde{\mathcal C}_{\overrightarrow{\mu}} (\lambda) 
                   e^{\frac{\sqrt{-1}}{2}\lambda\sum_{a=1}^3\kappa(\mu^{(a)})w_{a+1}/w_a}  
                   \prod_{a=1}^3 s^{(a)}_{\mu^{(a)}}, 
 \label{expansion_disconnected_G}
\end{eqnarray}
where 
\begin{align*}
  \kappa(\mu^{(a)}) 
      =\sum_{i=1}^\infty\mu^{(a)}_i(\mu^{(a)}_i-2i+1)
\end{align*} 
denotes the second Casimir invariant of $\mu^{(a)}$ and 
\begin{align*}
    s^{(a)}_{\mu^{(a)}}
        =\sum_{\nu\in\mathcal{P}}\frac{\chi_{\mu^{(a)}}(\nu)}{z_\nu}p^{(a)}_\nu. 
\end{align*}
$s^{(a)}_{\mu^{(a)}}$ is the Schur function $s_{\mu^{(a)}}$ 
obtained by substituting $p_\nu=p^{(a)}_\nu$. 

(\ref{expansion_disconnected_G}) shows that
$\exp G(\lambda;\overrightarrow{p};\overrightarrow{w})$ 
depends on $\overrightarrow{w}$ in a particular form.    
It is a consequence of the {\it invariance theorem} \cite{Li_Liu_Liu_Zhou_2009} 
for a generating function of formal relative Gromov-Witten invariants of 
$\mathbb{C}^3$.  
These invariants count stable maps 
from possibly disconnected curves to $\mathbb{C}^3$. 
In the localization computation of these invariants, 
the fixed points are labelled by a triple of partitions.  
The contribution from each fixed point takes the form of 
three-partition Hodge integrals multiplied by the double Hurwitz numbers of 
$\mathbb{P}^1$. 
Thus, the generating function of the formal relative 
Gromov-Witten invariants is a sum of these products over the sets of partitions. 
Though these terms depend on $\overrightarrow{w}$, 
the invariance theorem implies that the generating function 
itself does not depend on $\overrightarrow{w}$ in total.  
Accordingly, $\overrightarrow{w}$-dependence of each term 
should cancel out after the summation. 
With the aid of a combinatorial expression \cite{Liu_Liu_Zhou_2007} 
of the double Hurwitz numbers, 
the cancellation eventually yields the expansion (\ref{expansion_disconnected_G}). 

Three-partition Hodge integrals at special values of $\tau$ reduce to 
two-partition Hodge integrals since the $\psi$-classes can be renumbered 
in terms of two partitions at such values of $\tau$. 
For instance, at $\tau=1$,  
\begin{eqnarray}
   \begin{aligned}
     G_{g,(\mu^{(1)},\,\mu^{(2)},\,\mu^{(3)})} (1) 
       & = (-1)^{|\mu^{(1)}|-l(\mu^{(1)})} 
                   \frac{z_{\mu^{(1)}\cup\mu^{(2)}}}{z_{\mu^{(1)}}z_{\mu^{(2)}}} 
             G_{g,(\emptyset,\,\mu^{(1)}\cup\mu^{(2)},\,\mu^{(3)})} (1) 
     \\[0.2mm]
       & \qquad 
          +\delta_{g, 0}\sum_{m=1}^\infty \frac{(-1)^{m+1}}{m} 
                   \delta_{(\mu^{(1)},\,\mu^{(2)},\,\mu^{(3)}),\,(\emptyset,(m),(2m))}, 
   \end{aligned}
 \label{reduction}
\end{eqnarray}
where $\mu^{(1)}\cup\mu^{(2)}$ denotes 
the partition obtained by the union of the parts of $\mu^{(1)}$ and $\mu^{(2)}$ 
which are rearranged in descending order.  
The second term in the RHS is an anomalous contribution from the unstable cases.  
In terms of the generating functions (\ref{connected_G}), 
the reduction formula (\ref{reduction}) takes the form 
\begin{eqnarray}
   \begin{aligned}
    & \exp\left( 
               G\left(\lambda;(p^{(1)},p^{(2)},p^{(3)});1\right) 
               \right) \\[0.5mm]
    & \qquad 
         =\exp\left( G\left(\lambda;(0,p^+,p^{(3)});1\right) \right)
                   \exp\left( 
                         \sum_{m=1}^\infty\frac{(-1)^{m+1}}{m} p_m^{(1)}p_{2m}^{(3)}
                         \right), 
   \end{aligned}
  \label{reduction_formula_G}  
\end{eqnarray}
where $p^+ =(p^+_k)_{k=1}^\infty$, $p^+_k =(-1)^{k+1}p^{(1)}_k+p^{(2)}_k$. 

An explicit expression of the coefficients 
$\widetilde{\mathcal C}_{\overrightarrow{\mu}}(\lambda)$ 
can be read out from (\ref{reduction_formula_G}) 
by plugging (\ref{expansion_disconnected_G}) into (\ref{reduction_formula_G}): 
\begin{eqnarray}
  \begin{aligned}
   & e^{-\sqrt{-1}\lambda (-\kappa(\mu^{(1)})/2+\kappa(\mu^{(2)})+\kappa(\mu^{(3)})/4)}\, 
          \widetilde{\mathcal C}_{(\mu^{(1)},\,\mu^{(2)},\,\mu^{(3)})}\left(\lambda\right) 
   \\[2mm] 
   & \quad 
      = \sum_{\nu^1,\,\nu^3,\,\nu^+,\,\eta^1,\,\eta^3\,\in\mathcal{P}} 
               c_{\,\ltrans{\,\eta}^1\,\nu^1}^{\mu^{(1)}}\,
                      c_{\,\ltrans{\,\nu}^1\,\mu^{(2)}}^{\nu^+}\, 
                             c_{\,\eta^3 \,\nu^3}^{\mu^{(3)}}\, 
                e^{-\sqrt{-1}\lambda (\kappa(\nu^+)+\kappa(\nu^3)/4)}\, 
         \widetilde{\mathcal C}_{(\emptyset,\,\nu^+,\,\nu^3)}\left(\lambda\right)  
    \\
    & \qquad\qquad\qquad\qquad \times\, 
            \sum_{\xi\,\in\mathcal{P}} 
                       \frac{\chi_{\eta^1}(\xi) \chi_{\eta^3}(2\xi)}{z_\xi}, 
  \end{aligned}
 \label{reduction_formula_C}
\end{eqnarray}
where 
$2\xi =(2\xi_i)_{i=1}^\infty$ for $\xi=(\xi_i)_{i=1}^\infty$, 
and 
$c_{\,\ltrans{\,\eta}^1\,\nu^1}^{\mu^{(1)}}$, 
$c_{\,\ltrans{\,\nu}^1\,\mu^{(2)}}^{\nu^+}$ and 
$c_{\,\eta^3\,\nu^3}^{\mu^{(3)}}$ 
are the Littlewood-Richardson numbers (\ref{def_LR}). 
Furthermore, the {\it cut-and-join equations\,} 
for the two-partition Hodge integrals \cite{Liu_Liu_Zhou_2007} 
imply that 
\begin{align}
   \widetilde{\mathcal C}_{(\emptyset,\,\nu^+,\,\nu^3)}(\lambda)
     = q^{-\kappa(\nu^3)/2} 
        s_{\nu^+}(q^{-\rho}) 
        s_{\ltrans{\,\nu}^3}(q^{-\nu^+-\rho}), 
    \label{two-partition_C}
\end{align}
where $q=e^{-\sqrt{-1}\lambda}$, and 
$s_{\nu^+}(q^{-\rho})$ and $s_{\ltrans{\,\nu}^3}(q^{-\nu^+-\rho})$ 
are the special values of the infinite-variate Schur functions 
$s_{\nu^+}(\bm{x})$ and $s_{\ltrans{\,\nu}^3}(\bm{x})$ 
at $q^{-\rho}=\bigl(q^{i-1/2}\bigr)_{i=1}^\infty$ and 
$q^{-\nu^+-\rho}=\bigl(q^{-\nu^+_i+i-1/2}\bigr)_{i=1}^\infty$. 
As a consequence of (\ref{reduction_formula_C}) and (\ref{two-partition_C}), 
$\widetilde{\mathcal C}_{\overrightarrow{\mu}}(\lambda)$ 
can be expressed in a closed form \cite{Li_Liu_Liu_Zhou_2009}: 
\begin{eqnarray}
   \begin{aligned}
     & \widetilde{\mathcal C}_{\overrightarrow{\mu}} (\lambda)
          = q^{\kappa(\mu^{(1)})/2-\kappa(\mu^{(2)})-\kappa(\mu^{(3)})/4} 
     \\[1.5mm]
     & \qquad \times\!\!
              \sum_{\nu^1,\nu^3,\nu^+,\eta^1,\eta^3\,\in\mathcal{P}}\!\!
                    c_{\,\ltrans{\,\eta}^1\,\nu^1}^{~\mu^{(1)}}\,
                    c_{\,\ltrans{\,\nu}^1\,\mu^{(2)}}^{~\nu^+}\,
                    c_{\,\eta^3\,\ltrans{\,\nu}^3}^{~\mu^{(3)}}\,
                    q^{\kappa(\nu^+)+\kappa(\nu^3)/4} 
                    s_{\nu^+}(q^{-\rho})s_{\nu^3}(q^{-\nu^+-\rho}) 
     \\[0.2mm]
     & \qquad \qquad \qquad \qquad \times 
              \sum_{\xi\,\in\mathcal{P}} 
                   \frac{\chi_{\eta^1}(\xi)\chi_{\eta^3}(2\xi)}{z_\xi}. 
   \end{aligned}
 \label{LLLZ_formula}
\end{eqnarray}

\subsection{Topological vertex and three-partition Hodge integrals}

The toric data of a non-compact toric Calabi-Yau threefold 
are encoded in the associated fan of rational cones of dimension $\leq 3$ 
on $\mathbb{R}^3$. A plane section of this fan yields 
a triangulated polyhedron.  Its dual graph is trivalent, 
and referred to as the ``web'' or ``toric'' diagram.  
Each vertex of this trivalent graph is given a vertex weight 
called the topological vertex. 
According to the proposal of Aganagic et al. \cite{AKMV}, 
the topological string amplitudes on the Calabi-Yau threefold 
can be obtained by gluing these vertex weights along the edges of the graph.

\begin{figure}[tb] 
  \centering\includegraphics[scale=0.37]{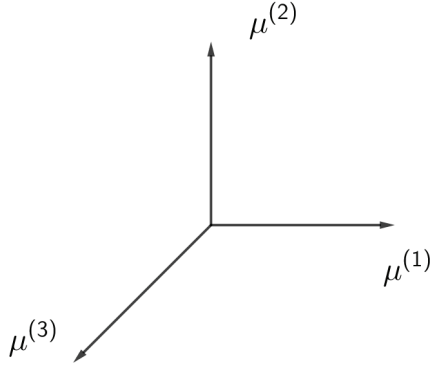}
  \caption{\small A diagrammatic representation of the topological vertex 
               (\ref{topological_vertex_def}).}   
  \label{figure_top_vertex}
\end{figure}

Let $q$ be a parameter in the range $0 <|q|<1$. 
The vertex weight at each vertex is labelled by a triple of partitions 
$\overrightarrow{\mu} 
=(\mu^{(1)},\mu^{(2)},\mu^{(3)})\in\mathcal{P}^3$, 
and defined as\footnote{ 
We follow a definition used in the recent literature 
\cite{Iqbal-Kzcaz-Vafa, Taki}. 
This definition differs from the earlier one \cite{AKMV, ADKMV} 
in that $q$ is replaced by $q^{-1}$ and an overall factor of the form 
$q^{\sum_{a=1}^3\kappa(\mu^{(a)})/2}$ is multiplied.
} 
\begin{eqnarray}
   \mathcal{C}_{\overrightarrow{\mu}}(q)
      = q^{\kappa(\mu^{(1)})/2}s_{\ltrans{\,\mu}^{(2)}}(q^{-\rho})
          \sum_{\eta\,\in\mathcal{P}}
             s_{\mu^{(1)}/\eta}(q ^{-\ltrans{\,\mu}^{(2)}-\rho}) 
             s_{\ltrans{\,\mu}^{(3)}/\eta}(q ^{-\mu^{(2)}-\rho}). 
  \label{topological_vertex_def}
\end{eqnarray}
$\mu^{(1)},\mu^{(2)}$ and $\mu^{(3)}$ are assigned to the three legs of 
the trivalent vertex numbered in a counterclockwise direction. 
See Figure \ref{figure_top_vertex}.
$s_{\mu^{(1)}/\eta}(q^{-\ltrans{\,\mu}^{(2)}-\rho})$ 
and $s_{\ltrans{\,\mu}^{(3)}/\eta}(q ^{-\mu^{(2)}-\rho})$ 
are the special values of the infinite-variate skew Schur functions 
$s_{\mu^{(1)}/\eta} (\bm{x})$ and $s_{\ltrans{\,\mu}^{(3)}/\eta}(\bm{x})$ 
at $q ^{-\ltrans{\,\mu}^{(2)}-\rho} =(q^{-\ltrans{\,\mu}^{(2)}_i+i-1/2})_{i=1}^\infty$ 
and $q ^{-\mu^{(2)}-\rho} =(q^{-\mu^{(2)}_i +i-1/2})_{i=1}^\infty$. 

We introduce a generating function of the topological vertex. Let 
$\bm{x}=\left(x_i\right)_{i=1}^\infty$, $\bm{y}=\left(y_i\right)_{i=1}^\infty$ and 
$\bm{z}=\left(z_i\right)_{i=1}^\infty$ be three sets of infinitely many variables. 
Define the generating function as 
\begin{align}
   \mathcal{W}\left(q;\bm{x},\bm{y},\bm{z};\overrightarrow{w}\right) 
       = \sum_{\overrightarrow{\mu}\in \mathcal{P}^3} 
                   \mathcal{C}_{\overrightarrow{\mu}}\left(q\right) 
                   q^{-\frac{1}{2}\sum_{a=1}^3\kappa(\mu^{(a)})(1+w_{a+1}/w_a)} 
                   s_{\mu^{(1)}}(\bm{x}) s_{\mu^{(2)}}(\bm{y}) s_{\mu^{(3)}}(\bm{z}).  
 \label{generating_function_topological_vertex} 
\end{align} 
In the case where $\overrightarrow{w}$ takes the form (\ref{tau}), 
we use the abbreviated notation 
\begin{align*}
    \mathcal{W}\left(q;\bm{x},\bm{y},\bm{z};\tau\right) 
         = \mathcal{W}\left(q;\bm{x},\bm{y},\bm{z};1,\tau,-1-\tau\right). 
\end{align*}

\bigskip

Our goal is to show the equivalence between 
the aforementioned generating functions of 
the topological vertex and three-partition Hodge integrals. 
More precisely, we prove the following: 
\begin{theorem} 
\label{theorem_conjecture} 
Let $q=e^{-\sqrt{-1}\lambda}$. 
The generating function (\ref{generating_function_topological_vertex}) 
coincides with the exponential of the generating function (\ref{connected_G}) as 
\begin{eqnarray}
    \mathcal{W}\left(q;\bm{x},\bm{y},\bm{z};\overrightarrow{w}\right) 
       = \exp\left(G\left(\lambda;\overrightarrow{p};\overrightarrow{w}\right)\right), 
 \label{eq1_theorem_conjecture}
\end{eqnarray} 
where the variables $\bm{x}$, $\bm{y}$ and $\bm{z}$ are related to 
$\overrightarrow{p}=(p^{(1)},p^{(2)},p^{(3)})$ as 
\begin{align}
    p^{(1)}_k =p_k(\bm{x}),\quad p^{(2)}_k=p_k(\bm{y}), \quad 
    p^{(3)}_k =p_k(\bm{z}), \quad 
    k=1,2,\dots ~.
\label{identify_p_x}
\end{align} 
In other words, 
the coefficients $\widetilde{\mathcal{C}}_{\overrightarrow{\mu}}(\lambda)$ 
of (\ref{expansion_disconnected_G}) are connected with the topological vertex as   
\begin{align}
     \widetilde{\mathcal{C}}_{\overrightarrow{\mu}}(\lambda)
            = q^{-\frac{1}{2}\sum_{a=1}^3\kappa(\mu^{(a)})} 
                    \mathcal{C}_{\overrightarrow{\mu}}(q), \quad 
     \forall\,\overrightarrow{\mu}=(\mu^{(1)},\mu^{(2)},\mu^{(3)})\in\mathcal{P}^3. 
  \label{tildeC=C}
\end{align}
\end{theorem} 

\bigskip

An immediate corollary of this result is cyclic symmetry of the topological vertex, 
which means that it is invariant under a cyclic permutation of three partitions 
assigned to the three legs, i.e., 
\begin{align}
    \mathcal{C}_{(\mu^{(1)},\mu^{(2)},\mu^{(3)})}(q) 
        =\mathcal{C}_{(\mu^{(2)},\mu^{(3)},\mu^{(1)})}(q), \quad 
    \forall\,\overrightarrow{\mu} =(\mu^{(1)},\mu^{(2)},\mu^{(3)})\in\mathcal{P}^3. 
 \label{cyclicity_topological_vertex}
\end{align}
Actually, three-partition Hodge integral is easily found to be invariant 
under simultaneous cyclic permutations of 
$\overrightarrow{\mu} =(\mu^{(1)},\mu^{(2)},\mu^{(3)})$ 
and $\overrightarrow{w}=(w_1,w_2,w_3)$ as
\begin{align*}
   G_{g,\,\mu^{(1)},\mu^{(2)},\mu^{(3)}}(w_1,w_2,w_3)
        = G_{g,\,\mu^{(2)},\mu^{(3)},\mu^{(1)}}(w_2,w_3,w_1). 
\end{align*}
This fact implies the cyclic symmetry of 
$\mathcal{C}_{\overrightarrow{\mu}}(q)$ by (\ref{tildeC=C}). 

\subsection{Integrable structure at positive integer values of $\tau$}

The reduction formula (\ref{reduction_formula_G}) 
of three-partition Hodge integrals can be extended to the case 
where $\tau$ takes any positive integer value $N = 1,2,\ldots$.  
Moreover, this leads to an unexpected relation with 
the generalized KdV hierarchies.  

The generalized KdV hierarchies are reduced systems of the KP hierarchy 
(see Dickey's book \cite{Dickey_book} for details of these integrable hierarchies). 
The KP hierarchy in the Lax formalism is the system 
\begin{align*}
  \frac{\partial L}{\partial t_k} = [B_k,L], \quad k = 1,2,\ldots,
\end{align*}
of Lax equations for the pseudo-differential operator 
\begin{align*}
   L = \partial_x +\sum_{n=1}^\infty u_{n+1}\partial_x^{-n}, \quad 
     \partial_x = \frac{\partial}{\partial x}, 
\end{align*}
where the coefficients $u_n$ are functions of the time variables 
$\bst = (t_1,t_2,\ldots$), $t_1$ is identified with $x$, 
and $B_k$ is the differential operator part 
\begin{align*}
  B_k = (L^k)_{+} = \partial_x^k + b_{k2}\partial_x^{k-2} + \cdots + b_{kk}
\end{align*}
of the $k$-th power of $L$.  
If the $N+1$-st power $\calL = L^{N+1}$ of $L$ is a differential operator, i.e., 
\begin{align*}
  \calL = B_{N+1} = \partial_x^{N+1} + b_2\partial_x^{N-1} + \cdots + b_{N+1}, 
\end{align*}
$L$ does not depend on $t_{m(N+1)}$, $m = 1,2,\dots$: 
\begin{equation*}
   \frac{\partial L}{\partial t_{m(N+1)}} = 0, \quad 
      m = 1,2,\ldots. 
\end{equation*}
This condition conversely characterizes the condition $L^{N+1} = B_{N+1}$.  
The remaining Lax equations can be reduced to the Lax equations 
\begin{align*}
  \frac{\partial\calL}{\partial t_k} = [B_k,\calL], \quad k = 1,2,\ldots, 
\end{align*}
for $\calL$.  These Lax equations comprises 
the $N$-th generalized KdV hierarchy.  
The KdV hierarchy amounts to the case where $N = 1$. 

This reduction procedure can be reformulated in terms of 
the tau function $\calT = \calT(\bst)$. 
The tau function yields the dressing operator 
\begin{align*}
  W = 1 + \sum_{n=1}^\infty w_n\partial_x^{-n} 
\end{align*}
via the generating function
\begin{equation*}
  w(\bst,z) 
    = 1 + \sum_{n=1}^\infty w_nz^{-n} 
    = \frac{\calT(t_1-z^{-1},\ldots,t_k-z^{-k}/k,\ldots)}
         {\calT(t_1,\ldots,t_k,\ldots)}. 
\end{equation*}
The dressing operator, in turn, defines the Lax operator as 
\begin{align*}
  L = W\partial_xW^{-1}.
\end{align*}
Consequently, if the tau function satisfies the equations 
\begin{equation}
  \frac{\partial^2\log\calT}{\partial t_k\partial t_{m(N+1)}} = 0, \quad 
  k,m = 1,2,\ldots,
  \label{red-cond2} 
\end{equation}
the coefficients of $W$ and $L$ do not depend 
on $t_{m(N+1)}$, $m = 1,2,\ldots$.  
This is a sufficient condition for the KP hierarchy 
to reduce to the $N$-th generalized KdV hierarchy.  
Note that this reduction condition is weaker than 
the commonly used condition 
\begin{equation*}
  \frac{\partial\calT}{\partial t_{m(N+1)}} = 0, \quad 
  m = 1,2,\ldots. 
\end{equation*}
It is the weaker condition (\ref{red-cond2}) that is relevant 
to the generating function (\ref{expansion_disconnected_G}) 
of three-partition Hodge integrals. 

Let $\calT(\lambda,N,p^{(1)},p^{(2)},\bst)$ 
denote the function of $\bst = (t_k)_{k=1}^\infty$ 
obtained from the generating function (\ref{expansion_disconnected_G})  
by substituting $t_k = p^{(3)}_k/k$, $k = 1,2,\ldots$: 
\begin{equation}
   \calT(\lambda,N,p^{(1)},p^{(2)},\bst) 
      = \exp\left(G(\lambda;\,\vec{p};\,N)\right). 
\label{def_calT}  
\end{equation}
$\lambda,N,p^{(1)}$ and $p^{(2)}$ are treated as parameters.  

\begin{theorem} 
\label{theorem_NKdV}
Let $N$ be a positive integer. 
$\calT(\lambda,N,p^{(1)},p^{(2)},\bst)$ is a tau function 
of the KP hierarchy that satisfies the condition (\ref{red-cond2}).  
The associated Lax operator $L$ satisfies the reduction condition 
$L^{N+1} = B_{N+1}$ for the $N$-th generalized KdV hierarchy.  \\
\end{theorem} 

\subsubsection*{\underline{Organization of the article}} 

We start Section 2 
with a brief review on a two-dimensional charged free fermion system. 
Various operators on the fermionic Fock space, including 
a realization of the two-dimensional quantum torus algebra, are introduced. 
These operators are used as fundamental tools 
in the discussions through the article.  
In Proposition \ref{fermionic_formula_W(q;x,y,z;tau)}, 
we give a fermionic representation of the generating function 
of the topological vertex. 

In Section 3, 
we explain the generalized shift symmetries. 
These symmetries act on a set of basis elements $V_m^{(k)}$ of 
the quantum torus algebra so as to shift the indices $k, m$ in a certain way. 
The shift symmeries are formulated by special values 
of the vertex operators $\Gamma_\pm(x)$ and $\Gamma'_\pm(x)$ 
and the exponential operator $q^{K/2}=\exp(\frac{K}{2}\log q)$, as
\begin{align*}
    V^{(k)}_m\mathbb{L}_\emptyset = (-1)^k\mathbb{L}_\emptyset V^{(k)}_{m-k}, \quad  
    V^{(-k)}_m\mathbb{L}'_\emptyset = \mathbb{L}'_\emptyset V^{(-k)}_{m-k}, \quad 
    q^{K/2}V_m^{(k)}q^{-K/2} =V_m^{(k-m)}, 
\end{align*}
where $\mathbb{L}_\emptyset =\Gamma_+ (q^{-\rho})\Gamma_-(q^{-\rho})$ and 
$\mathbb{L}'_\emptyset =\Gamma'_+ (q^{-\rho})\Gamma'_-(q^{-\rho})$ 
are the products of multi-variate vertex operators 
$\Gamma_\pm(\bm{x}) =\prod_{i=1}^\infty \Gamma_\pm(x_i)$ and 
$\Gamma'_\pm(\bm{x})=\prod_{i=1}^\infty \Gamma'_\pm(x_i)$ specialized to $q^{-\rho}$. 
These shift symmetries can be generalized by replacing 
$\mathbb{L}_\emptyset$ and $\mathbb{L}'_\emptyset$ with 
\begin{align*}
  \mathbb{L}_\alpha =s_\alpha(q^{-\rho})
                                \Gamma_-(q^{-\alpha-\rho})
                                \Gamma_+(q^{-\ltrans{\,\alpha}-\rho}), \quad 
  \mathbb{L}'_\alpha =s_\alpha(q^{-\rho})
                                \Gamma'_-(q^{-\alpha-\rho})
                                \Gamma'_+(q^{-\ltrans{\,\alpha}-\rho}) 
\end{align*}
and modifying $q^{K/2}$ in a certain manner. 
The generalized shift symmetries are summarized 
in Propositions \ref{ext_shift_symmetry_quote_1} 
and \ref{ext_shift_symmetry_quote_2}.  

The fermionic representation of the generating function 
in Proposition \ref{fermionic_formula_W(q;x,y,z;tau)} 
involves a weighted sum of operators of the form 
\begin{align*}
    \sum_{\alpha \in \mathcal{P}}\mathbb{G}_\alpha(\tau)s_\alpha, 
\end{align*}
where 
$\displaystyle 
   \mathbb{G}_\alpha(\tau) 
     = q^{-\kappa(\alpha)/2} q^{-\tau K/2} \mathbb{L}_\alpha q^{\tau K/2(1+\tau)}$ 
is a particular linear combination of the generalized shift symmetries. 
The generalized shift symmetries imply that, at certain values of $\tau$, 
the foregoing sum is factorized into a triple product of operators. 
The factorization formula at $\tau=1$ is given in 
Theorem \ref{factorization_formulas_tau=1}. 

In Section 4, 
we prove Theorem \ref{theorem_conjecture} 
using the results obtained in the preceding section. 
The aforementioned fermionic representation 
of the generating function of the topological vertex 
can be converted into a new representation 
by Theorem \ref{factorization_formulas_tau=1}. 
The resulting representation is presented  
in Proposition \ref{fermionic_formula_W(q;x,y,z;1)}. 
With the aid of the new representation of the generating function, 
we can reconsider its Schur function expansion and 
eventually find out the relation (\ref{tildeC=C}). 
Theorem \ref{theorem_conjecture} is thus proved. 

In Section 5, 
we prove Theorem \ref{theorem_NKdV}. 
The foregoing weighted sum of operators can be factorized in the cases 
where $\tau =1/N$, $N=2,3,\dots$, as well. 
The factorized expression is presented in Theorem \ref{factorization_formulas_1/N}. 
This formula yields in Proposition \ref{fermionic_formula_W(q;x,y,z;N)} 
a new representation of the generating functions of the topological vertex 
at positive integral values of $\tau$. 
Theorem \ref{theorem_conjecture} and 
Proposition \ref{fermionic_formula_W(q;x,y,z;N)} lead to a reduction formula of 
the generating functions of three-partition Hodge integrals 
in Proposition \ref{reduction_formula_G_tau=N}. 
This formula gives a generalization of (\ref{reduction}) 
at all positive integral values of $\tau$. 
We then consider integrable hierarchies underlying the generating functions 
of two-partition Hodge integrals. By combining these considerations, 
we eventually obtain Theorem \ref{theorem_NKdV}. 

\subsubsection*{\bf \underline{Acknowledgements}}
This work is supported by JSPS KAKENHI 
Grant Numbers JP15K04912 and JP18K03350.

\section{Fermionic representations of generating functions}

\subsection{Fermionic Fock space and operators}

Let $\psi_n,\psi^*_n,n\in\mathbb{Z}$, denote the Fourier modes of 
two-dimensional charged free fermion fields 
$$
  \psi (z) =\sum_{n\in\mathbb{Z}}\psi_nz^{-n-1},\quad 
  \psi^*(z) =\sum_{n\in\mathbb{Z}}\psi^*_nz^{-n}.
$$  
They satisfy the anti-commutation relations 
$$
  \psi_m\psi^*_n+\psi^*_n\psi_m =\delta_{m+n,0},\quad 
  \psi_m\psi_n+\psi_n\psi_m =0,\quad \psi^*_m\psi^*_n+\psi^*_n\psi^*_m =0. 
$$

The associated Fock space and its dual space are decomposed into 
the charge-$p$ sector for $p\in\mathbb{Z}$. 
Only the charge-$0$ sector is relevant to the subsequent computations. 
An orthonormal basis of the charge-$0$ sector is given by the ground states 
\begin{align*}
  & \langle0| =\langle-\infty|\dots\psi^*_{-i+1}\dots\psi^*_{-1}\psi^*_0,  
\\[2mm]
  & |0\rangle =\psi_0\psi_1\dots\psi_{i-1}\dots|-\infty\rangle 
\end{align*}
and the excited states 
\begin{align*}
  & \langle\lambda| 
       =\langle-\infty| \dots 
            \psi^*_{\lambda_i -i+1}\dots\psi^*_{\lambda_2-1}\psi^*_{\lambda_1}, 
\\[2mm] 
  & |\lambda\rangle 
       =\psi_{-\lambda_1}\psi_{-\lambda_2+1}\dots\psi_{-\lambda_i+i-1} 
            \dots|-\infty\rangle    
\end{align*}
labelled by partitions. The normal ordered product is prescribed by 
\begin{align*}
  :\psi_m\psi^*_n:\,\,=\,\psi_m\psi^*_n-\langle 0|\psi_m\psi^*_n|0\rangle.
\end{align*}

Fundamental tools in our computations are 
the following operators on the Fock space which preserve the charge. 
\begin{enumerate}

\item[(i)] 
The zero modes 
$$
    L_0 = \sum_{n\in\mathbb{Z}}n:\psi_{-n}\psi^*_n:,\quad 
    W_0 = \sum_{n\in\mathbb{Z}}n^2:\psi_{-n}\psi^*_n: 
$$
of the Virasoro and $\mbox{W}_3$ algebras and the Fourier modes 
$$
  J_m = \sum_{n\in\mathbb{Z}}:\psi_{m-n}\psi^*_{n}:,\quad 
  m\in\mathbb{Z}
$$
of the $U(1)$ current 
$J(z) = \sum_{m\in\mathbb{Z}}J_mz^{-m-1}=:\psi (z)\psi^* (z):$. 

\item[(ii)] 
The fermionic realization 
$$
  K = W_0-L_0+J_0/4 
     = \sum_{n\in\mathbb{Z}}(n-1/2)^2:\psi_{-n}\psi^*_n:
$$
of the so called ``cut-and-join operator'' \cite{Goulden_Jackson_1997}. 

\item[(iii)] 
The basis elements 
$$
  V^{(k)}_m 
   =q^{-k(m+1)/2}\sum_{n\in\mathbb{Z}}q^{kn}:\psi_{m-n}\psi^*_n: 
        - \,\frac{q^{k/2}}{1-q^k}\,\delta_{m,0}, \quad 
              k\in\mathbb{Z}_{\neq 0},~ m\in\mathbb{Z} 
$$
and 
$$
  V^{(0)}_m =J_m, \quad m\in\mathbb{Z} 
$$
of a fermionic realization\footnote{ 
The normalization of $V^{(k)}_m$ for $k\neq 0$ differs from 
\cite{Nakatsu_Takasaki_CMP_2009, Nakatsu_Takasaki_ASPS} 
in that an overall factor $q^{-k/2}$ is multiplied and that a constant term
$q^{k/2}(1-q^k)^{-1}\delta_{m,0}$ is subtracted.} 
of the quantum torus algebra, 
which satisfy the commutation relations 
$$
  \left[\,V^{(k)}_m,\,V^{(l)}_n\,\right] 
   = \left(q^{-\frac{kn-lm}{2}}-q^{\frac{kn-lm}{2}}\right) 
        V^{(k+l)}_{m+n}+m\,\delta_{k+l,0}\,\delta_{m+n,0}, \quad 
            k,l,m,n\in\mathbb{Z}.
$$

\item[(iv)] 
The vertex operators 
$$
  \Gamma_\pm(z) 
        = \exp\left(\sum_{k=1}^\infty\frac{z^k}{k}J_{\pm k}\right), \quad  
  \Gamma'_\pm(z) 
       = \exp\left(-\sum_{k=1}^\infty\frac{(-z)^k}{k}J_{\pm k}\right)
$$
and the multi-variable extensions 
$$
  \Gamma_\pm(\bm{x}) 
       = \prod_{i=1}^\infty\Gamma_{\pm}(x_i), \quad 
  \Gamma'_\pm(\bm{x}) 
      = \prod_{i=1}^\infty\Gamma'_{\pm}(x_i). 
$$
\end{enumerate}

The matrix elements of those operators are well-known. 
$J_0=V_0^{(0)}$, $L_0$ and $W_0$ are diagonal to the basis 
$\left\{|\lambda\rangle\right\}_{\lambda\in\mathcal{P}}$ 
in the charge-$0$ sector as
$$
  \langle\lambda|J_0|\mu\rangle 
          = 0,\quad 
  \langle\lambda|L_0|\mu\rangle 
          = \delta_{\lambda\mu}|\lambda|,\quad  
  \langle\lambda|W_0|\mu\rangle 
          = \delta_{\lambda\mu}(\kappa(\lambda)+|\lambda|). 
$$
Consequently, we have
\begin{align}
  \langle\lambda|K|\mu\rangle 
          = \delta_{\lambda\mu}\kappa(\lambda). 
\label{matrix_element_K}
\end{align}
Other zero modes of the quantum torus algebra are 
likewise diagonal to the basis 
$\left\{|\lambda\rangle\right\}_{\lambda\in\mathcal{P}}$. 
Their matrix elements are of the form 
$$
   \langle\lambda|V^{(k)}_0|\mu\rangle 
          = \delta_{\lambda\mu}\varphi_k(\mu), 
   \quad k\in\mathbb{Z}_{\neq 0}, 
$$
where
\begin{align*}
  \varphi_k(\mu) 
    =\left\{\begin{array}{rl} 
             \displaystyle 
             -\sum_{i=1}^\infty q^{k\left(-\ltrans{\,\mu}_i+i-1/2\right)}
   & \mbox{for $k\geq 1$}, \\[2mm]
             \displaystyle 
               \sum_{i=1}^\infty q^{-k\left(-\mu_i+i-1/2\right)} 
   & \mbox{for $k\leq -1$}. 
   \end{array}\right.
\end{align*}

It is also well known that the matrix elements of 
$\Gamma_\pm(\bm{x})$ and $\Gamma'_\pm(\bm{x})$ 
are skew Schur functions 
\begin{align}
  & \langle\lambda|\Gamma_-(\bm{x})|\mu\rangle 
      = \langle\mu|\Gamma_+(\bm{x})|\lambda\rangle =s_{\lambda /\mu}(\bm{x}),
  \label{matrix_element_Gamma} 
  \\[2.5mm]
  & \langle\lambda|\Gamma'_-(\bm{x})|\mu\rangle 
      = \langle\mu|\Gamma'_+(\bm{x})|\lambda\rangle 
      = s_{\,\ltrans{\,\lambda}/\ltrans{\,\mu}}(\bm{x}). 
  \label{matrix_element_Gamma'}
\end{align}

\subsection{Fermionic representation of generating function 
(\ref{generating_function_topological_vertex})}

Let us provide a fermionic representation of the generating function 
(\ref{generating_function_topological_vertex}). 
We first introduce a set of operator-valued functions labeled by partitions as  
\begin{align}
   \mathbb{G}_\alpha (\tau) 
     = q^{-\frac{\kappa(\alpha)}{2\tau}}s_\alpha (q^{-\rho})\,
          q^{-\frac{\tau}{2}K}\Gamma_-\left(q^{-\alpha -\rho}\right)
             \Gamma_+\bigl(q^{-\ltrans{\,\alpha}-\rho}\bigr)\,
                q^{\frac{\tau}{2(1+\tau)}K},\quad 
   \alpha\in\mathcal{P},~ \tau\neq 0,-1, 
    \label{G_alpha(tau)} 
\end{align}
where $q^{-\tau K/2}$ and $q^{\tau K/2(1+\tau)}$ 
are the exponential operators 
$\exp\bigl(-\frac{\tau K}{2}\log q\bigr)$, 
$\exp\bigl(\frac{\tau K}{2(1+\tau)}\log q\bigr)$, 
and $\Gamma_-\left(q^{-\alpha -\rho}\right)$ 
and $\Gamma_+\bigl(q^{-\ltrans{\,\alpha}-\rho}\bigr)$ 
denote respectively the multi-variate operators 
$\Gamma_-(\bm{x})$ and $\Gamma_+(\bm{x})$ 
specialized at $q^{-\alpha -\rho}$ and $q^{-\ltrans{\,\alpha}-\rho}$. 
The matrix element of (\ref{G_alpha(tau)}) yields 
the topological vertex (\ref{topological_vertex_def}) as follows:
\begin{proposition} 
\label{prop_fermion_expression_topological_vertex}
\begin{eqnarray}
     \mathcal{C}_{\overrightarrow{\mu}}(q)
       = q^{ \frac{1+\tau}{2}\kappa (\mu^{(1)})
              -\frac{1}{2\tau}\kappa(\mu^{(2)}) 
              +\frac{\tau}{2(1+\tau)}\kappa(\mu^{(3)})} 
          \langle\mu^{(1)}| 
                \mathbb{G}_{\,\ltrans{\,\mu}^{(2)}}(\tau) 
          |\ltrans{\,\mu}^{(3)}\rangle, 
      \label{fermion_expression_topological_vertex}
\end{eqnarray} 
where $\overrightarrow{\mu}=(\mu^{(1)},\mu^{(2)},\mu^{(3)})$. 
\end{proposition} 

\paragraph{\it Proof\/.}
We compute the matrix element 
in the RHS of (\ref{fermion_expression_topological_vertex})  
by interposing the identity
 $\sum_{\eta\in\mathcal{P}}|\eta\rangle\langle\eta|=1$ 
in the charge-$0$ sector as follows: 
\begin{align}
   & \langle\mu^{(1)}|
           \mathbb{G}_{\,\ltrans{\,\mu}^{(2)}}(\tau) 
                |\ltrans{\,\mu}^{(3)}\rangle 
   \nonumber \\
   & = q^{-\frac{\kappa(\ltrans{\,\mu}^{(2)})}{2\tau}} 
                s_{\,\ltrans{\,\mu}^{(2)}}(q^{-\rho})
          \langle\mu^{(1)}|
                 q^{-\frac{\tau}{2}K}\,
                 \Gamma_-(q^{-\ltrans{\,\mu}^{(2)}-\rho})
                       \left(\sum_{\eta\in\mathcal{P}}|\eta\rangle\langle\eta|\right)
                 \Gamma_+(q^{-\mu^{(2)}-\rho})
                 q^{\frac{\tau}{2(1+\tau)}K} 
         |\ltrans{\,\mu}^{(3)}\rangle 
   \nonumber \\[1mm]
   & =q^{\frac{\kappa(\mu^{(2)})}{2\tau}}s_{\,\ltrans{\,\mu}^{(2)}}(q^{-\rho})
          \sum_{\eta\in\mathcal{P}}
          \langle\mu^{(1)}| 
               q^{-\frac{\tau}{2}K}\,\Gamma_-(q^{-\ltrans{\,\mu}^{(2)}-\rho}) 
          |\eta\rangle \langle\eta| 
               \Gamma_+(q^{-\mu^{(2)}-\rho})q^{\frac{\tau}{2(1+\tau)}K}
          |\ltrans{\,\mu}^{(3)}\rangle. 
  \label{fermion_expression_topological_vertex_proof}
\end{align}
By (\ref{matrix_element_K}) and (\ref{matrix_element_Gamma}), 
the matrix elements in the RHS can be expressed as 
\begin{align*}
      \langle\mu^{(1)}|
          q^{-\frac{\tau}{2}K}\,\Gamma_-(q^{-\ltrans{\,\mu}^{(2)}-\rho})
               |\eta\rangle 
      & = q^{-\frac{\tau}{2}\kappa(\mu^{(1)})}
              s_{\mu^{(1)}/\eta}(q^{-\ltrans{\,\mu}^{(2)}-\rho}), 
\\[2.5mm] 
      \langle\eta|
            \Gamma_+(q^{-\mu^{(2)}-\rho})
              q^{\frac{\tau}{2(1+\tau)}K}
                   |\ltrans{\,\mu}^{(3)}\rangle 
      & = q^{-\frac{\tau}{2(1+\tau)}\kappa(\mu^{(3)})}
              s_{\,\ltrans{\,\mu}^{(3)}/\eta}(q^{-\mu^{(2)}-\rho}). 
\end{align*} 
By plugging those expressions into 
(\ref{fermion_expression_topological_vertex_proof}), we find    
\begin{align*}
  & \langle\mu^{(1)}| 
         \mathbb{G}_{\,\ltrans{\,\mu}^{(2)}}(\tau) 
            |\ltrans{\,\mu}^{(3)}\rangle 
\\[1mm]
  & \quad 
     = q^{-\frac{\tau}{2}\kappa(\mu^{(1)})+\frac{1}{2\tau}\kappa(\mu^{(2)}) 
           -\frac{\tau}{2(1+\tau)}\kappa(\mu^{(3)})} 
           s_{\ltrans{\,\mu}^{(2)}}(q^{-\rho})  
               \sum_{\eta \in \mathcal{P}}
                  s_{\mu^{(1)}/\eta}(q^{-\ltrans{\,\mu}^{(2)}-\rho})
                  s_{\,\ltrans{\,\mu}^{(3)}/\eta}(q^{-\mu^{(2)}-\rho}) 
\\
  & \quad 
     = q^{-\frac{1+\tau}{2}\kappa(\mu^{(1)})+\frac{1}{2\tau}\kappa(\mu^{(2)})
            -\frac{\tau}{2(1+\tau)}\kappa(\mu^{(3)})} 
        \mathcal{C}_{\overrightarrow{\mu}}(q). 
\end{align*}
Thus we obtain (\ref{fermion_expression_topological_vertex}). 
\qed \\

The generating function (\ref{generating_function_topological_vertex})  
has a fermionic expression of the following form due to 
(\ref{fermion_expression_topological_vertex}): 
\begin{proposition} 
\label{fermionic_formula_W(q;x,y,z;tau)}
\begin{eqnarray}
    \mathcal{W}(q;\bm{x},\bm{y},\bm{z};\tau)
       = \langle 0|
            \Gamma_+(\bm{x})
             \biggl(
                   \sum_{\alpha\in\mathcal{P}}
                   \mathbb{G}_\alpha (\tau) 
                   s_{\,\ltrans{\,\alpha}}(\bm{y})
             \biggr) 
          \Gamma'_-(\bm{z})|0\rangle. 
     \label{W(q;x,y,z;tau)_fermion_expression}
\end{eqnarray} 
\end{proposition} 

\noindent
\paragraph{\it Proof\/.} 
Using (\ref{fermion_expression_topological_vertex}), 
three sums in the RHS of (\ref{generating_function_topological_vertex}) 
for $\overrightarrow{w}=(1,\tau,-1-\tau)$ become 
\begin{align*}
   \mathcal{W}(q;\bm{x},\bm{y},\bm{z};\tau) 
     & = \sum_{\overrightarrow{\mu}\in\mathcal{P}^3} 
          \langle\mu^{(1)}|
              \mathbb{G}_{\,\ltrans{\,\mu}^{(2)}}(\tau) 
                 |\ltrans{\,\mu}^{(3)}\rangle
                   s_{\mu^{(1)}}(\bm{x})s_{\mu^{(2)}}(\bm{y})s_{\mu^{(3)}}(\bm{z})
     \nonumber \\[1mm]
     & = \Bigl( 
                    \sum_{\mu^{(1)}\in\mathcal{P}}s_{\mu^{(1)}}(\bm{x}) 
                          \langle\mu^{(1)}|
            \Bigr) 
       \sum_{\mu^{(2)}\in\mathcal{P}} 
                   \mathbb{G}_{\,\ltrans{\,\mu}^{(2)}}(\tau)s_{\mu^{(2)}}(\bm{y}) 
            \Bigl(
               \sum_{\mu^{(3)}\in\mathcal{P}}
                |\ltrans{\,\mu}^{(3)}\rangle s_{\mu^{(3)}}(\bm{z}) 
            \Bigr) 
     \nonumber \\[1mm]
     & = \langle 0|\Gamma_+(\bm{x})
           \biggl( 
                \sum_{\alpha\in\mathcal{P}} 
                    \mathbb{G}_{\alpha}(\tau)s_{\,\ltrans{\,\alpha}}(\bm{y}) 
           \biggr)
         \Gamma'_-(\bm{z})|0\rangle, 
\end{align*} 
where (\ref{matrix_element_Gamma}) and (\ref{matrix_element_Gamma'}) 
are used in the last line. 
\qed 

\section{Generalized shift symmetry and factorization formula}

The shift symmetries 
\cite{Nakatsu_Takasaki_CMP_2009, Nakatsu_Takasaki_ASPS} 
act on the basis elements of the quantum torus algebra. 
There are three different types of shift symmetries: 
\begin{itemize}
\item[(i)]  
For $k\geq 1$ and $m\in\mathbb{Z}$, 
\begin{eqnarray}
     V^{(k)}_m\,\Gamma_-(q^{-\rho})\Gamma_+(q^{-\rho})
        = (-1)^k\Gamma_-(q^{-\rho}) \Gamma_+(q^{-\rho})\,V^{(k)}_{m-k}. 
     \label{shift_symmetry_1}
\end{eqnarray}

\item[(ii)] 
For $k\geq 1$ and $m\in\mathbb{Z}$, 
\begin{eqnarray}
     V^{(-k)}_{m}\,\Gamma'_-(q^{-\rho})\Gamma'_+(q^{-\rho})
         =\Gamma'_-(q^{-\rho})\Gamma'_+(q^{-\rho})\,V^{(-k)}_{m-k}. 
      \label{shift_symmetry_2}
\end{eqnarray}

\item[(iii)] 
For $k,m\in\mathbb{Z}$,  
\begin{eqnarray}
     q^{K/2}V^{(k)}_mq^{-K/2}=V^{(k-m)}_m. 
       \label{shift_symmetry_3}
\end{eqnarray}
\end{itemize}

These symmetries together with their combinations are used to clarify 
the integrable structures of various melting crystal models 
\cite{Nakatsu_Takasaki_CMP_2009, Nakatsu_Takasaki_ASPS}  
and open topological string amplitudes 
\cite{Takasaki_Nakatsu_JPA_2016}, \cite{Takasaki_2014}. 
Likewise, quantum curves of open topological strings 
on the closed topological vertex and the strip geometries 
are derived \cite{Takasaki_Nakatsu_SIGMA_2017} using the symmetries. 
An extension of the symmetries is investigated in the companion paper 
\cite{Nakatsu_Takasaki_2019}. 
The generalized shift symmetries obtained therein provides a powerful tool 
to prove Theorem \ref{theorem_conjecture}. 

\subsection{Generalized shift symmetries on quantum torus algebra}

\begin{figure}[t] 
   \centering\includegraphics[scale=0.48]{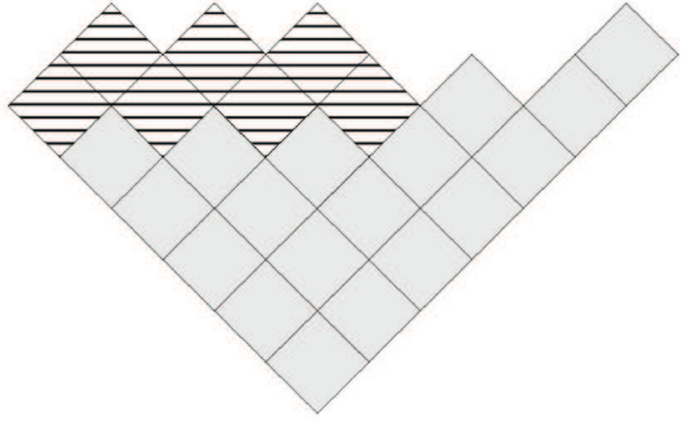}
   \caption{\small 
        The Young diagrams $\alpha =(7,5,4,4,3,2)$ and $\beta=(7,5,3,2,1)$ 
        are laid to overlap. 
        The skew Young diagram $\alpha/\beta$ is indicated with hatched boxes 
        and is a ribbon of length $7$ and height $3$.}   
    \label{figure_1}
\end{figure}

The skew Young diagram $\alpha/\beta$ is called a ribbon 
iff there exist positive integers $r$ and $s$, $r>s$, such that partitions 
$\alpha=(\alpha_i)_{i=1}^\infty$ and $\beta=(\beta_i)_{i=1}^\infty$ 
correlate to each other through the condition
\begin{align*}
     \bigl\{\alpha_i-i\bigr\}_{i=1}^N\cup\left\{\beta_r-r\right\}
       =\bigl\{\beta_i-i\bigr\}_{i=1}^N\cup\left\{\alpha_s-s\right\}, \quad 
       \forall\,N>r, 
\end{align*}
where 
$\bigl\{\alpha_i-i\bigr\}_{i=1}^N$ and $\bigl\{\beta_i-i\bigr\}_{i=1}^N$ 
are finite subsets of $\mathbb{Z}$ consisting of $N$ distinct integers. 
See Figure \ref{figure_1}.  
The semi-infinite subsets 
$\bigl\{\alpha_i-i\bigr\}_{i=1}^\infty$ and $\bigl\{\beta_i-i\bigr\}_{i=1}^\infty$ 
are referred to the Maya diagrams of $\alpha$ and $\beta$.  
The length of a ribbon $\alpha/\beta$ is given by $|\alpha|-|\beta|$ and the height, 
denoted by ${\sf ht}(\alpha/\beta)$, is given by $r-s$. 

The set of partitions of which subtraction from $\alpha$ leaves 
a ribbon of length $k$ is denoted by 
\begin{align*} 
    \mathcal{R}_{k,\alpha}^{(-)} 
       =\Bigl.\Bigl\{\beta\in\mathcal{P}\,\Bigr|\, 
              \mbox{$\alpha/\beta$ is a ribbon of length $k$.}\Bigr\}.  
\end{align*}
Likewise, the set of partitions which leave a ribbon of length $k$ 
by subtraction of $\alpha$ is 
\begin{align*} 
   \mathcal{R}_{k,\alpha}^{(+)} 
        =\Bigl.\Bigl\{\beta\in\mathcal{P}\,\Bigr|\, 
                \mbox{$\beta/\alpha$ is a ribbon of length $k$.}\Bigr\}. 
\end{align*}
We describe these sets in a single symbol indexed by 
$\alpha\in\mathcal{P}$ and $k\in\mathbb{Z}$, $k\neq 0$, as 
\begin{align*}
    \mathcal{R}_{k,\alpha} 
    =\left\{\begin{array}{ll} 
         \displaystyle \mathcal{R}_{k,\alpha}^{(-)}
                                & \quad \mbox{for $k\geq 1$}, 
         \\[2mm]
        \displaystyle \mathcal{R}_{-k,\alpha}^{(+)} 
                                & \quad \mbox{for $k\leq -1$}.
        \end{array}\right. 
\end{align*}
That means that $\mathcal{R}_{k,\alpha}$ consists of partitions 
which can be obtained from $\alpha$ 
by removing a ribbon of length $k$ for $k\geq 1$ or 
by adding a ribbon of length $-k$ for $k\leq -1$. 

For $\alpha\in\mathcal{P}$ and $\beta\in\mathcal{R}_{k,\alpha}$, 
we introduce their relative sign by  
\begin{align*}
   {\sf sgn}(\alpha,\beta) 
   =\left\{\begin{array}{ll}  
           (-1)^{{\sf ht}(\alpha/\beta)} 
                      & \quad \mbox{for $k\geq 1$}, 
           \\[2mm] 
           (-1)^{{\sf ht}(\beta/\alpha)} 
                      & \quad \mbox{for $k\leq -1$}. 
              \end{array}\right.
\end{align*}
Since $\beta\in\mathcal{R}_{k,\alpha}$ implies 
$\alpha\in\mathcal{R}_{-k,\beta}$, 
the relative sign is symmetric under the exchange of partitions 
$$
    {\sf sgn}(\alpha,\beta) = {\sf sgn}(\beta,\alpha).
$$

\subsubsection{Generalized shift symmetries} 

Let us explain the extension of the shift symmetries (i), (ii) and (iii) 
which is derived in \cite{Nakatsu_Takasaki_2019}. 
For $\alpha\in\mathcal{P}$, we introduce an operator of the form  
\begin{align}
  \mathbb{L}_\alpha 
         = s_\alpha (q^{-\rho}) 
                \Gamma_-\left(q^{-\alpha-\rho}\right) 
                     \Gamma_+\bigl(q^{-\ltrans{\,\alpha}-\rho}\bigr).  
    \label{L_alpha} 
\end{align} 
Note that (\ref{shift_symmetry_1}) is written 
with the aid of $\mathbb{L}_\emptyset$ as 
$$
    V^{(k)}_m\,\mathbb{L}_\emptyset 
              = (-1)^k\mathbb{L}_\emptyset\,V^{(k)}_{m-k},\quad 
    k\geq 1,~m\in\mathbb{Z}. 
$$
We generalize the shift symmetry (i) by replacing 
$\mathbb{L}_\emptyset$ in the LHS with $\mathbb{L}_\alpha$. 
This yields an extension of the following form.
\begin{proposition} 
\label{ext_shift_symmetry_quote_1}
{\sf \cite{Nakatsu_Takasaki_2019}} 
For $k\in\mathbb{Z}$, $k\neq 0$ and $m\in\mathbb{Z}$, 
we have 
\begin{eqnarray} 
    V^{(k)}_m\,\mathbb{L}_\alpha 
         = (-1)^k\,\mathbb{L}_\alpha\,V^{(k)}_{m-k} 
            + q^{k J_0} 
                 \sum_{\beta\in\mathcal{R}_{k,\alpha}} 
                     \mbox{\sf sgn}(\alpha,\beta) 
                          q^{-m(\kappa(\alpha)-\kappa(\beta))/2k}
                               \,\mathbb{L}_\beta. 
\label{ext_shift_symmetry_1}
\end{eqnarray} 
As regards the Heisenberg sub-algebra generated by $V^{(0)}_m$, we have  
\begin{eqnarray}
     V^{(0)}_m\mathbb{L}_\alpha 
         = \mathbb{L}_\alpha V^{(0)}_m 
              +\varphi_{-m}(\alpha)\mathbb{L}_\alpha. 
    \label{ext_shift_symmetry_1_sub}
\end{eqnarray}
\end{proposition} 

\begin{remark} 
(\ref{ext_shift_symmetry_1}) of 
$\alpha =\emptyset$ with $k\geq 1$ reproduces (\ref{shift_symmetry_1}). 
Otherwise, there is no counterpart in (i). For instance, 
$\mathcal{R}_{-l,\emptyset}=\mathcal{R}^{(+)}_{l,\emptyset}$ $(l \geq 1)$ 
consists of  hooks $(1^{l-j}j^1)$ of weight $l$. Thus, rewriting $k=-l$, 
(\ref{ext_shift_symmetry_1}) of $\alpha =\emptyset$ with $k\leq -1$ reads  
\begin{align*} 
  V^{(-l)}_m\mathbb{L}_\emptyset  
     = (-1)^l\mathbb{L}_\emptyset V^{(-l)}_{m+l} 
        + q^{-lJ_0}
             \sum_{j=1}^{l} 
                 (-1)^{l-j}q^{m(l+1-2j)/2}
                      \mathbb{L}_{(1^{l-j}j^1)}, 
  \quad \mbox{$l\geq 1$.}
\end{align*}
\end{remark} 

\bigskip

The shift symmetry (ii) can be generalized in a manner similar to 
the case of (i) as described in Proposition \ref{ext_shift_symmetry_quote_1}. 
With $\alpha\in\mathcal{P}$, we associate an operator of the form  
\begin{align}
  \mathbb{L}'_\alpha = s_\alpha(q^{-\rho}) 
        \Gamma'_-\left(q^{-\alpha-\rho}\right) 
        \Gamma'_+\bigl(q^{-\ltrans{\,\alpha}-\rho}\bigr). 
     \label{L'_alpha}  
\end{align} 
Note that (\ref{shift_symmetry_2}) is written 
with the aid of $\mathbb{L}'_\emptyset$ as 
$$
   V^{(-k)}_m\,\mathbb{L}_\emptyset 
       = (-1)^k\mathbb{L}_\emptyset\,V^{(-k)}_{m-k}, 
    \quad k\geq 1,~m\in\mathbb{Z}. 
$$
The symmetry (ii) can be generalized 
by replacing $\mathbb{L}'_\emptyset$ in the LHS 
with $\mathbb{L}'_\alpha$. 
We then obtain an extension of the following form. 
\begin{proposition} 
\label{ext_shift_symmetry_quote_2}
{\sf \cite{Nakatsu_Takasaki_2019}}
For $k\in\mathbb{Z}$, $k\neq 0$ and $m\in\mathbb{Z}$, 
we have 
\begin{eqnarray} 
   V^{(-k)}_m\mathbb{L}'_\alpha 
      = \mathbb{L}'_\alpha V^{(-k)}_{m-k} 
         +(-1)^{m+1} q^{-kJ_0} 
             \sum_{\beta\in\mathcal{R}_{k,\alpha}} 
                \mbox{\sf sgn}(\alpha,\beta) 
                   q^{-m\left(\kappa(\alpha)-\kappa(\beta)\right)/2k}
                     \,\mathbb{L}'_\beta.  
      \label{ext_shift_symmetry_2} 
\end{eqnarray}
As regards the generators of the Heisenberg sub-algebra, we have 
\begin{eqnarray}
   V^{(0)}_m\mathbb{L}'_\alpha 
       =\mathbb{L}'_\alpha V^{(0)}_m 
             +(-1)^{m+1} 
                  \varphi_{-m}(\alpha)\mathbb{L}'_\alpha. 
   \label{ext_shift_symmetry_2_sub}
\end{eqnarray}
\end{proposition} 

\begin{remark} 
(\ref{ext_shift_symmetry_2}) of $\alpha =\emptyset$ with $k\geq 1$ 
reproduces (\ref{shift_symmetry_2}). 
Otherwise, there is no counterpart in (ii). 
\end{remark} 

The shift symmetry (iii) can be generalized as follows: 
\begin{proposition} 
\label{ext_shift_symmetry_quote_3} 
For $k,m\in\mathbb{Z}$, we have 
\begin{eqnarray}
    q^{\gamma K}V^{(k)}_m q^{-\gamma K} 
          = V^{\left(k-2m\gamma\right)}_m,  
     \label{ext_shift_symmetry_3}
\end{eqnarray}
where $\gamma$ is chosen to be a rational number 
which satisfies $2m\gamma\in\mathbb{Z}$. 
\end{proposition} 

\paragraph{\it Proof\/.}
The commutation relations 
\begin{align*}
   \left[K,\psi_n\right] =\bigl(n+\frac{1}{2}\bigr)^2\psi_n, \quad 
   \left[K,\psi^*_n\right] =-\bigl(n-\frac{1}{2}\bigr)^2\psi^*_n
\end{align*}
are integrated to give 
\begin{align}
   q^{\gamma K}\psi_nq^{-\gamma K} 
             =q^{\gamma\left(n+1/2\right)^2}\psi_n, \quad 
   q^{\gamma K}\psi^*_nq^{-\gamma K} 
             =q^{-\gamma\left(n-1/2\right)^2}\psi_n. 
 \label{generalized_shift_symmetry_3_proof}
\end{align}
Consider the case of $m\neq 0$. 
Using (\ref{generalized_shift_symmetry_3_proof}), 
the LHS of (\ref{ext_shift_symmetry_3}) is computed as 
\begin{align*}
  q^{\gamma K}V^{(k)}_mq^{-\gamma K}  
     & =q^{-k(m+1)/2}\sum_nq^{kn} 
               q^{\gamma K}:\psi_{m-n}\psi^*_n:q^{-\gamma K}  \\ 
     & =q^{-k(m+1)/2}\sum_nq^{kn+\gamma\left(n-n+1/2\right)^2 
                  -\gamma\left(n-1/2\right)^2}:\psi_{m-n}\psi^*_n:  \\
     & =q^{-\left(k-2m\gamma\right)\left(m+1\right)/2} 
                \sum_nq^{\left(k-2m\gamma\right)n}:\psi_{m-n}\psi^*_n: \\ 
     & =V^{(k-2m\gamma)}_m.
\end{align*}
In the case of $m=0$, since $K$ commutes with $V^{(k)}_0$,  
we have $q^{\gamma K}V^{(k)}_0q^{-\gamma K}=V^{(k)}_0$. 
Thus, we obtain (\ref{ext_shift_symmetry_3}). 
\qed 

\subsubsection{Particular combinations of generalized shift symmetries}

Combining (\ref{ext_shift_symmetry_1}) and (\ref{ext_shift_symmetry_3}), 
we obtain an operator-valued identity.    
\begin{proposition} 
\label{ext_shift_symmetry_A_tau}  
For $m\in\mathbb{Z}$, $m\neq 0$,  
at values of $\tau$ satisfying $m\tau\in\mathbb{Z}$ and $m\tau\neq 0$,  
we have the operator-valued identity  
\begin{eqnarray}
   J_{-m}\mathbb{G}_\alpha (\tau) 
      =\mathbb{G}_\alpha (\tau)(-1)^{m\tau}J_{-m(1+\tau)} 
         +q^{m\tau J_0} 
             \sum_{\beta\in\mathcal{R}_{m\tau,\alpha}}  
                \mbox{\sf sgn}(\alpha,\beta) 
                   \mathbb{G}_\beta(\tau).  
     \label{ext_shift_symmetry_a_tau}
\end{eqnarray}
\end{proposition} 

\paragraph{\it Proof\/.} 
Using $\mathbb{L}_\alpha$, 
we express $\mathbb{G}_\alpha (\tau)$ in (\ref{G_alpha(tau)}) as 
$$
    \mathbb{G}_\alpha (\tau)
        =q^{-\kappa(\alpha)/2\tau}q^{-\tau K/2} 
            \mathbb{L}_\alpha q^{\tau K/2(1+\tau)}.
$$ 
The symmetry (\ref{ext_shift_symmetry_3}) with $\gamma =\tau/2$ 
reads 
$$
     q^{\tau K/2}J_{-m}q^{-\tau K/2} =V_{-m}^{(m\tau)}, 
$$
which converts $J_{-m}\mathbb{G}_\alpha (\tau)$ to 
\begin{align}
     J_{-m}\mathbb{G}_\alpha (\tau) 
        & =q^{-\kappa(\alpha)/2\tau} 
                  J_{-m}q^{-\tau K/2} 
                     \mathbb{L}_\alpha q^{\tau K/2(1+\tau)}  
        \nonumber \\[2mm]
        & =q^{-\kappa(\alpha)/2\tau} 
               q^{-\tau K/2}V_{-m}^{(m\tau)} 
                   \mathbb{L}_\alpha q^{\tau K/2(1+\tau)}. 
        \label{proof_symmetry_A_tau_1}
\end{align} 
Using (\ref{ext_shift_symmetry_1}), 
the RHS of (\ref{proof_symmetry_A_tau_1}) is further converted to  
\begin{align}
     & \mbox{RHS of (\ref{proof_symmetry_A_tau_1})} 
     \nonumber \\[1mm]
     & =q^{-\kappa(\alpha)/2\tau} 
               q^{-\tau K/2}  
               \biggl\{ 
                     (-1)^{m\tau}\mathbb{L}_\alpha V_{-m(1+\tau)}^{(m\tau)} 
                        +q^{m\tau J_0} 
                            \sum_{\beta\in\mathcal{R}_{m\tau,\alpha}}
                                {\sf sgn}(\alpha,\beta) 
                                q^{\frac{\kappa(\alpha)-\kappa(\beta)}{2\tau}}  
                                \mathbb{L}_\beta
                \biggr\}
           q^{\tau K/2(1+\tau)}  
     \nonumber \\[2mm]
     &  =\mathbb{G}_\alpha (\tau) 
              (-1)^{m\tau}q^{-\tau K/2(1+\tau)} 
                   V_{-m(1+\tau)}^{(m\tau)}
                        q^{\tau K/2(1+\tau)} 
         +q^{m\tau J_0} 
           \sum_{\beta\in\mathcal{R}_{m\tau,\alpha}}
                 {\sf sgn}(\alpha,\beta) \mathbb{G}_\beta (\tau). 
         \label{proof_symmetry_A_tau_2}
\end{align}
Here, the symmetry (\ref{ext_shift_symmetry_3}) gives  
$$
   q^{-\tau K/2(1+\tau)}V_{-m(1+\tau)}^{(m\tau)}q^{\tau K/2(1+\tau)} 
        =V_{-m(1+\tau)}^{(0)} 
        =J_{-m(1+\tau)}.
$$ 
Thus, the last line in (\ref{proof_symmetry_A_tau_2}) becomes   
$$
    \mbox{RHS of (\ref{proof_symmetry_A_tau_2})} 
         =\mathbb{G}_\alpha (\tau) 
               (-1)^{m\tau}J_{-m(1+\tau)} 
           +q^{m\tau J_0} 
             \sum_{\beta \in \mathcal{R}_{m\tau,\alpha}} 
                   {\sf sgn}(\alpha,\beta)\mathbb{G}_\beta (\tau). 
$$ 
We therefore obtain (\ref{ext_shift_symmetry_a_tau}).  
\qed \\ 

Operator-valued identity analogous to (\ref{ext_shift_symmetry_a_tau}) 
can be obtained if we focus on (\ref{ext_shift_symmetry_2}) 
instead of (\ref{ext_shift_symmetry_1}). 
With $\alpha\in\mathcal{P}$, 
let us introduce an operator-valued function of the form 
\begin{align}
   \mathbb{G}'_\alpha(\tau) 
      = q^{-\kappa(\alpha)/2\tau}q^{\tau K/2} 
            \mathbb{L}'_\alpha 
                q^{-\tau K/2(1+\tau)},\quad 
    \tau\neq 0,-1.
    \label{G'_alpha(tau)} 
\end{align}
We note that the topological vertex (\ref{topological_vertex_def}) 
can be written in terms of the matrix element of (\ref{G'_alpha(tau)}) as 
\begin{align*}
  \mathcal{C}_{\overrightarrow{\mu}}(q) 
      =q^{\frac{1+\tau}{2}\kappa (\mu^{(1)}) 
              -\frac{1}{2\tau}\kappa(\mu^{(2)}) 
                   +\frac{\tau}{2(1+\tau)}\kappa(\mu^{(3)})}  
        \langle\ltrans{\,\mu}^{(1)}|
               \mathbb{G}'_{\,\ltrans{\,\mu}^{(2)}}(\tau) 
        |\mu^{(3)}\rangle. 
\end{align*}

\begin{proposition} 
\label{ext_shift_symmetry_B_tau} 
For $m\in\mathbb{Z}$, $m\neq 0$, 
at values of $\tau$ satisfying $m\tau\in\mathbb{Z}$, $m\tau\neq 0$,  
we have the operator-valued identity 
\begin{eqnarray} 
     J_{-m}\mathbb{G}'_\alpha(\tau) 
        =\mathbb{G}'_\alpha (\tau)J_{-m(1+\tau)}  
            +(-1)^{m+1}q^{-m\tau J_0} 
                \sum_{\beta\in\mathcal{R}_{m\tau,\alpha}} 
                   \mbox{\sf sgn}(\alpha,\beta)
                   \mathbb{G}'_\beta(\tau).  
        \label{ext_shift_symmetry_b_tau}
\end{eqnarray}
\end{proposition} 

\paragraph{\it Proof\/.}  
Using (\ref{ext_shift_symmetry_2}) instead of (\ref{ext_shift_symmetry_1}),     
the same computations as in the proof of Proposition \ref{ext_shift_symmetry_A_tau} 
give rise to (\ref{ext_shift_symmetry_b_tau}). 
\qed \\

\subsection{Factorization formulas at $\tau =1$}

We consider generating functions of 
the operator-valued functions (\ref{G_alpha(tau)}) and (\ref{G'_alpha(tau)}).  
These generating functions can be factorized at $\tau=1$.
Let us introduce infinite-variate functions  
\begin{align*}
    s_\alpha[\bm{t}] 
       =\langle 0|
            \exp\left(\sum_{k=1}^\infty t_kJ_k\right) 
         |\alpha\rangle, \quad  
    \bm{t}=(t_k)_{k=1}^\infty.
\end{align*}
It is well known that the functions $s_\alpha[\bm{t}]$ are converted to 
the Schur functions $s_\alpha(\bm{x})$ 
by the Miwa transformation 
\begin{align*}
    t_k = \frac{1}{k}\sum_{i=1}^\infty x_i^k,\quad k=1,2,\ldots\,, 
\end{align*}
and the Schur functions $s_{\ltrans{\,\alpha}}(\bm{x})$ 
by another version of the transformation 
\begin{align*}
    t_k = -\frac{1}{k}\sum_{i=1}^\infty (-x_i)^k, \quad k=1,2,\dots\,.
\end{align*}

We note that the partial derivatives of $s_\alpha[\bm{t}]$ 
with respect to $\bm{t}$ become as follows. 
\begin{proposition}  
\label{prop_partial_t_s[t]}  
\begin{eqnarray}
   \partial_{t_m}s_\alpha[\bm{t}] 
       =\sum_{\beta\in\mathcal{R}_{m,\alpha}^{(-)}} 
                  {\sf sgn}(\alpha,\beta) 
                  s_\beta[\bm{t}], \quad 
   m=1,2, \dots \,.
\label{partial_t_s[t]}
\end{eqnarray}
\end{proposition} 

\paragraph{\it Proof\/.}
We rewrite the partial derivatives as 
\begin{align}
   \partial_{t_m}s_\alpha[\bm{t}]  
      =\partial_{t_m} 
        \langle 0|\exp\left(\sum_{k=1}^\infty t_kJ_k\right)|\alpha\rangle 
      =\langle 0|\exp\left(\sum_{k=1}^\infty t_kJ_k\right)J_m|\alpha\rangle, 
     \label{proof_partial_t_s[t]}
\end{align}
which can be computed by use of a version of Murnaghan-Nakayama's rule: 
\begin{align*}
    J_k|\alpha\rangle 
       = \sum_{\beta\in\mathcal{R}_{k,\alpha}} 
           {\sf sgn}(\alpha,\beta)|\beta\rangle, \quad  
     k\in\mathbb{Z}_{\neq 0}.  
\end{align*} 
Applying the rule into the RHS of (\ref{proof_partial_t_s[t]}), 
we see   
\begin{align*}
   \mbox{RHS of (\ref{proof_partial_t_s[t]})} 
      =\sum_{\beta\in\mathcal{R}_{m,\alpha}^{(-)}} 
             {\sf sgn}(\alpha,\beta)  
             \langle 0| 
                  \exp\left(\sum_{k=1}^\infty t_kJ_k\right)
             |\beta\rangle 
     =\sum_{\beta\in\mathcal{R}_{m,\alpha}^{(-)}} 
            {\sf sgn}(\alpha,\beta)s_\beta[\bm{t}]. 
\end{align*}
Thus we obtain (\ref{partial_t_s[t]}). 
\qed

\subsubsection{Operator-valued generating functions}

We define operator-valued generating functions by 
\begin{align}
    \mathbb{G}[\bm{t};\tau] 
       & =\sum_{\alpha\in\mathcal{P}} 
              \mathbb{G}_\alpha (\tau)s_\alpha[\bm{t}], 
        \label{G[t;tau]} 
    \\[1mm] 
    \mathbb{G}'[\bm{t};\tau] 
       & =\sum_{\alpha\in\mathcal{P}} 
             \mathbb{G}'_\alpha (\tau)s_\alpha[\bm{t}].  
       \label{G'[t;tau]}
\end{align} 
\begin{proposition} 
\label{partial-differentials_G[t;tau]}
Partial derivatives of (\ref{G[t;tau]}) and (\ref{G'[t;tau]}) 
with respect to $\bm{t}=(t_k)_{k=1}^\infty$ 
are expressed in the following forms. 
\begin{align}
   \partial_{t_m}\mathbb{G}[\bm{t};\tau] 
      & =\sum_{\alpha\in\mathcal{P}} 
              \Biggl(
                  \sum_{\beta\in\mathcal{R}^{(+)}_{m,\alpha}} 
                    {\sf sgn} (\alpha,\beta)
                   \mathbb{G}_\beta(\tau)
              \Biggr) 
              s_\alpha[\bm{t}], 
         \label{partial_t_G[t;tau]} 
     \\[1mm]
  \partial_{t_m}\mathbb{G}'[\bm{t};\tau]  
      & =\sum_{\alpha\in\mathcal{P}}
              \Biggl( 
                 \sum_{\beta\in\mathcal{R}^{(+)}_{m,\alpha}} 
                    {\sf sgn}(\alpha,\beta) 
                    \mathbb{G}'_\beta(\tau) 
              \Biggr) 
              s_\alpha[\bm{t}], 
         \label{partial_t_G'[t;tau]}
\end{align} 
where $m=1,2,\dots$\,. 
\end{proposition} 

\paragraph{\it Proof\/.}
By (\ref{partial_t_s[t]}), 
the partial derivatives of $\mathbb{G}[\bm{t};\tau]$ read   
\begin{gather} 
   \partial_{t_m}\mathbb{G}[\bm{t};\tau] 
      =\sum_{\alpha\in\mathcal{P}} 
             \mathbb{G}_\alpha(\tau)\partial_{t_m}s_\alpha[\bm{t}] 
   \nonumber \\[1mm]
      =\sum_{\alpha\in\mathcal{P}} 
        \sum_{\beta\in\mathcal{R}^{(-)}_{m,\alpha}} 
             {\sf sgn}(\alpha,\beta) 
             \mathbb{G}_\alpha(\tau)s_\beta[\bm{t}] 
      =\sum_{\alpha\in\mathcal{P}}
        \sum_{\beta\in\mathcal{R}^{(+)}_{m,\alpha}} 
             {\sf sgn}(\beta,\alpha) 
             \mathbb{G}_\beta(\tau)s_\alpha[\bm{t}], 
    \label{proof_partial_t_G[t]}
\end{gather}
where the roles of $\alpha$ and $\beta$ are exchanged in the last line. 
Similarly, the partial derivatives of $\mathbb{G}'[\bm{t};\tau]$ 
can be expressed as 
\begin{align}
   \partial_{t_m}\mathbb{G}'[\bm{t};\tau] 
       =\sum_{\alpha\in\mathcal{P}} 
              \mathbb{G}'_\alpha (\tau)\partial_{t_m}s_\alpha[\bm{t}] 
       =\sum_{\alpha\in\mathcal{P}}
         \sum_{\beta\in\mathcal{R}^{(+)}_{m,\alpha}} 
              {\sf sgn}(\beta,\alpha)
              \mathbb{G}'_\beta(\tau)s_\alpha[\bm{t}]. 
    \label{proof_partial_t_G'[t]}
\end{align}
Taking account of the symmetry 
${\sf sgn}(\beta,\alpha)={\sf sgn}(\alpha,\beta)$,  
we find that (\ref{proof_partial_t_G[t]}) and (\ref{proof_partial_t_G'[t]}) 
yield (\ref{partial_t_G[t;tau]}) and (\ref{partial_t_G'[t;tau]}) respectively.  
\qed

\subsubsection{Factorization formulas at $\tau =1$}

The operator-valued generating functions 
$\mathbb{G}[\bm{t};\tau]$ and $\mathbb{G}'[\bm{t};\tau]$ at $\tau=1$ 
can be factorized into a triple product of operators. 
\begin{theorem} 
{\sf(factorization formulas at $\tau=1$)}
\label{factorization_formulas_tau=1} 
$\mathbb{G}[\bm{t};\tau]$ and $\mathbb{G}'[\bm{t};\tau]$ at $\tau=1$ 
satisfy the identities 
\begin{align}  
    \mathbb{G}[\bm{t};1] 
        & = \exp\left(\sum_{k=1}^\infty t_kq^{kJ_0}J_k\right) 
                    \mathbb{G}_\emptyset (1) 
             \exp\left(\sum_{k=1}^\infty (-1)^{k+1}t_kq^{kJ_0}J_{2k}\right),  
         \label{factorization_tau=1} 
  \\[2mm]
     \mathbb{G}'[\bm{t};1] 
        & = \exp\left(\sum_{k=1}^\infty (-1)^{k+1}t_kq^{-kJ_0}J_k\right)
                   \mathbb{G}'_\emptyset (1)
             \exp\left(\sum_{k=1}^\infty (-1)^kt_kq^{-kJ_0}J_{2k}\right), 
        \label{factorization'_tau=1}
\end{align}
where $\mathbb{G}_\emptyset (1)$ and $\mathbb{G}'_\emptyset (1)$ 
are operators of $\mathbb{G}_\emptyset (\tau)$ and 
$\mathbb{G}'_\emptyset (\tau)$ specialized to $\tau=1$.  
\end{theorem}  

\paragraph{\it Proof\/.}
Let us derive differential equations 
for $\mathbb{G}[\bm{t};1]$ and $\mathbb{G}'[\bm{t};1]$. 
We first consider the case of $\mathbb{G}\left[\bm{t};1\right]$. 
Partial derivatives of $\mathbb{G}\left[\bm{t};1\right]$ 
with respect to $\bm{t}$ are of the form (\ref{partial_t_G[t;tau]}).  
Note here that (\ref{ext_shift_symmetry_a_tau}) now reads 
\begin{align}
    J_{-k}\mathbb{G}_\alpha(1) 
       =\mathbb{G}_\alpha(1)(-1)^kJ_{-2k} 
           +q^{kJ_0}\sum_{\beta\in\mathcal{R}_{k,\alpha}}
               \mbox{\sf sgn}(\alpha,\beta)
               \mathbb{G}_\beta(1), 
      \quad k\in\mathbb{Z}_{\neq 0}.
   \label{ext_shift_symmetry_a1}
\end{align}
We rewrite (\ref{ext_shift_symmetry_a1}) for the cases of $k\leq -1$. 
By letting $k=-m$, these become   
\begin{align}
     \sum_{\beta\in\mathcal{R}^{(+)}_{m,\alpha}} 
             {\sf sgn}(\alpha,\beta)\mathbb{G}_\beta(1) 
        =q^{mJ_0}J_m\mathbb{G}_\alpha(1) 
              +\mathbb{G}_\alpha(1)(-1)^{m+1}q^{m J_0}J_{2m}, 
    \label{ext_shift_symmetry_a2}
\end{align}  
where $m=1,2,\dots$\,.  
By plugging (\ref{ext_shift_symmetry_a2}) into the double sum, 
the RHS of (\ref{partial_t_G[t;tau]}) turns out to be  
\begin{align*}
   \mbox{RHS of (\ref{partial_t_G[t;tau]}) at $\tau\!=\!1$} \quad  
       & =\sum_{\alpha\in\mathcal{P}} 
                 \Bigl(
                     q^{mJ_0}J_m\mathbb{G}_\alpha(1) 
                     +\mathbb{G}_\alpha(1)(-1)^{m+1}q^{mJ_0}J_{2m} 
                 \Bigr) 
             s_\alpha[\bm{t}] 
     \\[1mm]
       & =q^{mJ_0}J_m 
              \sum_{\alpha\in\mathcal{P}} 
                  \mathbb{G}_\alpha(1)s_\alpha\left[\bm{t}\right]  
            +\sum_{\alpha\in\mathcal{P}}
                  \mathbb{G}_\alpha(1)s_\alpha[\bm{t}] 
                   (-1)^{m+1}q^{mJ_0}J_{2m} 
      \\[1mm]
        & =q^{mJ_0}J_m\mathbb{G}[\bm{t};1] 
              +\mathbb{G}[\bm{t};1](-1)^{m+1}q^{mJ_0}J_{2m}. 
\end{align*} 
We thus find that $\mathbb{G}\left[\bm{t};1\right]$ satisfies 
the first-order differential equations  
\begin{align}
    \partial_{t_m}\mathbb{G}[\bm{t};1]
        =q^{mJ_0}J_m\mathbb{G}[\bm{t};1] 
          +\mathbb{G}[\bm{t};1](-1)^{m+1}q^{mJ_0}J_{2m}
  \label{PDE_G[t;1]}
\end{align} 
for $m=1,2,\dots$\,.

By the uniqueness of solutions of the initial value problem,  
(\ref{PDE_G[t;1]}) can be solved as 
\begin{align}
    \mathbb{G}[\bm{t};1] 
        =\exp\left(\sum_{k=1}^\infty t_kq^{kJ_0}J_k\right) 
                 \mathbb{G}[\bm{0};1] 
          \exp\left(\sum_{k=1}^\infty(-1)^{k+1}t_kq^{kJ_0}J_{2k}\right), 
   \label{GS_PDE_G[t;1]}
\end{align}
where $\mathbb{G}[\bm{0};1]$ denotes an operator at the initial time. 
Since $\mathbb{G}[\bm{0};1]=\mathbb{G}_\emptyset(1)$, 
(\ref{GS_PDE_G[t;1]}) yields (\ref{factorization_tau=1}).  

(\ref{factorization'_tau=1}) is likewise obtained by 
using Proposition \ref{ext_shift_symmetry_B_tau} 
in place of Proposition \ref{ext_shift_symmetry_A_tau}.  
Partial derivatives of $\mathbb{G}'[\bm{t};1]$ with respect to $\bm{t}$ 
are of the form (\ref{partial_t_G'[t;tau]}). 
Specialized to $\tau=1$, the double sum therein can be computed 
by Proposition \ref{ext_shift_symmetry_B_tau}. 
(\ref{ext_shift_symmetry_b_tau}) therein reads       
\begin{align}
   J_{-k}\mathbb{G}'_\alpha(1)
       =\mathbb{G}'_\alpha(1)J_{-2k} 
         +(-1)^{k+1}q^{-kJ_0} 
           \sum_{\beta\in\mathcal{R}_{k,\alpha}} 
                \mbox{\sf sgn}(\alpha,\beta) 
                \mathbb{G}'_\beta(1), 
   \quad k\in\mathbb{Z}_{\neq 0}. 
        \label{ext_shift_symmetry_b1}
\end{align} 
We rewrite (\ref{ext_shift_symmetry_b1}) for the cases of $k\leq -1$.  
By letting $k=-m$, these become    
\begin{align}
    \sum_{\beta\in\mathcal{R}^{(+)}_{m,\alpha}} 
    {\sf sgn}(\alpha,\beta)\mathbb{G}'_\beta(1) 
       =(-1)^{m+1} 
           \Bigl( 
                 q^{-mJ_0}J_m\mathbb{G}'_\alpha(1) 
                 -\mathbb{G}'_\alpha(1)q^{-m J_0}J_{2m} 
           \Bigr), 
  \label{ext_shift_symmetry_b2}
\end{align} 
where $m=1,2,\dots$\,. 
By plugging (\ref{ext_shift_symmetry_b2}) into the double sum, 
the RHS of (\ref{partial_t_G'[t;tau]}) is converted to   
\begin{align*}
    \mbox{RHS of (\ref{partial_t_G'[t;tau]}) at $\tau\!=\!1$} \quad 
    & = (-1)^{m+1} 
           \sum_{\alpha\in\mathcal{P}}
               \Bigl( 
                   q^{-mJ_0}J_m\mathbb{G}'_\alpha(1) 
                   -\mathbb{G}'_\alpha(1)q^{-mJ_0}J_{2m} 
               \Bigr) 
    \\[1mm]
    & = (-1)^{m+1}q^{-mJ_0}J_m\mathbb{G}'[\bm{t};1] 
           + \mathbb{G}'[\bm{t};1](-1)^mq^{-mJ_0}J_{2m}. 
\end{align*} 
We thus find that $\mathbb{G}'[\bm{t};1]$ satisfies 
the first-order differential equations   
\begin{align} 
   \partial_{t_m}\mathbb{G}'[\bm{t};1] 
        = (-1)^{m+1}q^{-mJ_0} J_m\mathbb{G}'[\bm{t};1] 
            + \mathbb{G}'[\bm{t};1](-1)^mq^{-mJ_0}J_{2m}
   \label{PDE_G'[t;1]}
\end{align}
for $m=1,2,\dots$\,.

General solutions of the differential equations (\ref{PDE_G'[t;1]}) 
are of the form 
\begin{align}
   \mathbb{G}'[\bm{t};1] 
      =\exp\left(\sum_{k=1}^\infty(-1)^{k+1}t_kq^{-kJ_0}J_k\right) 
               \mathbb{G}'[\bm{0};1] 
        \exp\left(\sum_{k=1}^\infty(-1)^kt_kq^{-kJ_0}J_{2k}\right), 
  \label{GS_PDE_G'[t;1]}
\end{align}
where $\mathbb{G}'[\bm{0};1]$ denotes an operator at the initial time. 
In the current case, 
we have $\mathbb{G}'[\bm{0};1]=\mathbb{G}'_\emptyset(1)$. 
Therefore, (\ref{GS_PDE_G'[t;1]}) yields (\ref{factorization'_tau=1}). 
\qed \\

\section{Proof of Theorem \ref{theorem_conjecture}} 

\subsection{Yet another representation at $\tau=1$}

In the case of $\tau=1$, 
another expression of generating function of the topological vertex 
can be derived from Theorem \ref{factorization_formulas_tau=1}. 
We conveniently start with an operator identity of the following form.
\begin{proposition} 
\begin{eqnarray}
  \mathbb{G}_\emptyset (1)\Gamma'_-(\bm{z}) 
      = \exp\left( 
                  \sum_{m=1}^\infty \frac{(-1)^{m+1}p_{2m}(\bm{z})}{m}J_{-m} 
              \right)  
         \mathbb{G}_{\emptyset} (1)\Gamma_-(\bm{z}), 
  \label{lemma_rewriting_W(q;x,y,z;1)}
\end{eqnarray}
\end{proposition} 

\paragraph{\it Proof\/.}
Let us rewrite $\Gamma'_-(\bm{z})$ as 
\begin{align}
  \Gamma'_-(\bm{z}) 
    & =\exp\left( 
             \sum_{m=1}^\infty\frac{p_{2m-1}(\bm{z})}{2m-1}J_{-2m+1}
             - \sum_{m=1}^\infty\frac{p_{2m}(\bm{z})}{2m}J_{-2m} 
             \right) 
  \nonumber \\[1mm]
    & =\exp\left( 
                  - \sum_{m=1}^\infty\frac{p_{2m}(\bm{z})}{m}J_{-2m}
                \right) 
         \Gamma_-(\bm{z}). 
    \label{proof_lemma_W(q;x,y,z;1)_1}
\end{align}
(\ref{ext_shift_symmetry_a1}) implies the relation 
\begin{align}
    J_{-m}\mathbb{G}_\emptyset (1) 
         = (-1)^m\mathbb{G}_\emptyset (1)J_{-2m}, \quad 
    m=1,2,\dots,\,.
    \label{proof_lemma_W(q;x,y,z;1)_2}
\end{align}
Combining (\ref{proof_lemma_W(q;x,y,z;1)_1}) 
with (\ref{proof_lemma_W(q;x,y,z;1)_2}), we see   
\begin{align*}
   \mathbb{G}_\emptyset(1)\Gamma'_-(\bm{z})
      & = \mathbb{G}_\emptyset(1)
                \exp\left( 
                    - \sum_{m=1}^\infty\frac{p_{2m}(\bm{z})}{m}J_{-2m} 
                       \right) 
            \Gamma_-(\bm{z})
     \\[1.5mm]
      & = \exp\left( 
                 \sum_{m=1}^\infty\frac{(-1)^{m+1}p_{2m}(\bm{z})}{m}J_{-m} 
                  \right)
            \mathbb{G}_\emptyset(1)\Gamma_-(\bm{z}).
\end{align*}
Thus we obtain (\ref{lemma_rewriting_W(q;x,y,z;1)}). 
\qed 

\bigskip

\begin{proposition} 
\label{fermionic_formula_W(q;x,y,z;1)}
The generating function (\ref{generating_function_topological_vertex}) 
at $\tau =1$ has a fermionic representation of the form  
\begin{align}
    \mathcal{W}\left(q;\bm{x},\bm{y},\bm{z};1\right)
        = \langle 0| 
              \Gamma_+(\bm{x})\Gamma'_+(\bm{y})
              \mathbb{G}_\emptyset (1)\Gamma_-(\bm{z})
           |0\rangle 
        \exp\left(
             \sum_{m=1}^\infty\frac{(-1)^{m+1}}{m}p_m(\bm{x})p_{2m}(\bm{z})
               \right). 
   \label{W(q;x,y,z;1)_fermion_expression}
\end{align}
\end{proposition} 

\paragraph{\it Proof\/.}   
We adopt the fermionic representation 
(\ref{W(q;x,y,z;tau)_fermion_expression}) for a description of 
the generating function (\ref{generating_function_topological_vertex}). 
Letting $\tau=1$, 
we rewrite (\ref{W(q;x,y,z;tau)_fermion_expression}) 
by using Theorem \ref{factorization_formulas_tau=1}. 
Actually, substituting $t_k=(-1)^{k+1}p_k(\bm{y})/k$ 
in (\ref{factorization_tau=1}), we find     
\begin{align*}
   \sum_{\alpha\in\mathcal{P}} 
          \mathbb{G}_\alpha (1)s_{\ltrans{\,\alpha}}(\bm{y}) 
      =\exp\left(
                 \sum_{k=1}^\infty
                      \frac{(-1)^{k+1}p_k(\bm{y})}{k}q^{kJ_0}J_k 
              \right)
         \mathbb{G}_\emptyset (1) 
         \exp\left(
                  \sum_{k=1}^\infty 
                      \frac{p_k(\bm{y})}{k}q^{kJ_0}J_{2k}
                \right).  
\end{align*} 
The sum of operators in (\ref{W(q;x,y,z;tau)_fermion_expression}) 
can be thus converted into a triple product of operators as 
\begin{align} 
   & \mathcal{W}(q;\bm{x},\bm{y},\bm{z};1)  
  \nonumber \\
   & =\langle 0|
             \Gamma_+(\bm{x}) 
               \exp\left(
                        \sum_{k=1}^\infty\frac{(-1)^{k+1}p_k(\bm{y})}{k}J_k
                      \right) 
             \mathbb{G}_\emptyset (1)  
                \exp\left(
                         \sum_{k=1}^\infty\frac{p_k(\bm{y})}{k}J_{2k}
                       \right) 
             \Gamma'_- (\bm{z})
        |0\rangle 
     \nonumber \\
     & =\langle 0| 
               \Gamma_+(\bm{x})\Gamma'_+(\bm{y})\mathbb{G}_\emptyset (1)
                   \exp\left(
                              \sum_{k=1}^\infty\frac{p_k(\bm{y})}{k}J_{2k}
                          \right)
              \Gamma'_-(\bm{z})
       |0\rangle.  
     \label{rewriting_W(q;x,y,z;1)_1}
\end{align}

We further rewrite the RHS of (\ref{rewriting_W(q;x,y,z;1)_1}). 
The commutation relations 
\begin{align*}
   \Gamma'_-(\bm{z})J_{2k} 
      =J_{2k}\Gamma'_-(\bm{z}) 
          +\Gamma'_-(\bm{z})p_{2k}(\bm{z}), 
\end{align*}
and the annihilation property 
$J_{2k}|0\rangle =0$ for $k\geq 1$ yield   
\begin{align} 
   \exp\left( 
             \sum_{k=1}^\infty\frac{p_k(\bm{y})}{k}J_{2k}
          \right)
   \Gamma'_-(\bm{z})|0\rangle 
   =\Gamma'_-(\bm{z})|0\rangle
   \exp\left( 
            -\sum_{k=1}^\infty\frac{p_k(\bm{y})p_{2k}(\bm{z})}{k}
          \right).  
     \label{rewriting_W(q;x,y,z;1)_2}
\end{align} 
By (\ref{rewriting_W(q;x,y,z;1)_2}), 
the RHS of (\ref{rewriting_W(q;x,y,z;1)_1}) can be converted to      
\begin{align}
   \mathcal{W}(q;\bm{x},\bm{y},\bm{z};1)
      =\langle 0|
           \Gamma_+(\bm{x})\Gamma'_+(\bm{y}) 
           \mathbb{G}_\emptyset(1)\Gamma'_-(\bm{z})
        |0\rangle 
           \exp\left( 
                  -\sum_{k=1}^\infty\frac{p_k(\bm{y})p_{2k}(\bm{z})}{k}
                  \right).
    \label{rewriting_W(q;x,y,z;1)_3}
\end{align}
The bra vector in (\ref{rewriting_W(q;x,y,z;1)_3}) can be expressed 
with the aid of (\ref{lemma_rewriting_W(q;x,y,z;1)}) as follows:    
\begin{eqnarray}
\begin{aligned}
   & \langle 0|
         \Gamma_+(\bm{x})\Gamma'_+(\bm{y})
         \mathbb{G}_\emptyset(1)\Gamma'_-(\bm{z})  
   \\[1mm]
   & =\langle 0| 
             \Gamma_+(\bm{x})\Gamma'_+(\bm{y}) 
             \exp\left(
                \sum_{m=1}^\infty\frac{(-1)^{m+1}p_{2m}(\bm{z})}{m}J_{-m}
                    \right) 
             \mathbb{G}_{\emptyset} (1)\Gamma_-(\bm{z}) 
   \\[1mm]
   & =\langle 0|
             \Gamma_+(\bm{x})\Gamma'_+(\bm{y}) 
             \mathbb{G}_{\emptyset} (1)\Gamma_-(\bm{z})
      \exp\biggl\{
              \sum_{m=1}^\infty
                  \frac{1}{m}\Bigl(
                          p_m(\bm{y})p_{2m}(\bm{z})
                          +(-1)^{m+1}p_m(\bm{x})p_{2m}(\bm{z})
                                 \Bigr)
              \biggr\}, 
\end{aligned}
\label{eq_rewriting_W(q;x,y,z;1)_3}
\end{eqnarray}
where the negative modes $J_{-m}$ have been moved to the left 
by the commutation relations
\begin{align*}
   \Gamma_+(\bm{x})J_{-m} 
   & =J_{-m}\Gamma_+(\bm{x}) 
         +\Gamma_+(\bm{x})p_{m}(\bm{x}),
   \\[2mm]
   \Gamma'_+(\bm{y})J_{-m}
   & =J_{-m}\Gamma'_+(\bm{y}) 
         +\Gamma'_+(\bm{y})(-1)^{m+1}p_{m}(\bm{y}), 
\end{align*} 
together  with $\langle 0|J_{-m}=0$ at the end. 
Designating the last part of (\ref{eq_rewriting_W(q;x,y,z;1)_3}) 
as the bra vector and its pairing with the ket vector $|0\rangle$ 
being substituted for the matrix element in (\ref{rewriting_W(q;x,y,z;1)_3}), 
we eventually obtain (\ref{W(q;x,y,z;1)_fermion_expression}).  
\qed \\

The matrix element in the RHS of (\ref{W(q;x,y,z;1)_fermion_expression}) 
has an expansion in terms of the Schur functions and the skew Schur functions.  
\begin{proposition} 
\label{proposition_Schur_expansion_fermion_rep}
The matrix element in the fermionic expression 
(\ref{W(q;x,y,z;1)_fermion_expression}) 
has a Schur function expansion of the form 
\begin{eqnarray}
\begin{aligned}
    & \langle 0|
          \Gamma_+(\bm{x})\Gamma'_+(\bm{y}) 
          \mathbb{G}_\emptyset (1)\Gamma_-(\bm{z})
       |0\rangle 
  \\[1mm]
    & =\sum_{\nu^1,\nu^3,\nu^+\in\mathcal{P}} 
       s_{\nu^1}(\bm{x}) 
       s_{\nu^+/\ltrans{\,\nu}^1}(\bm{y}) 
       s_{\,\ltrans{\,\nu}^3}(\bm{z}) 
       q^{\kappa(\nu^+)+\kappa(\nu^3)/4}
       s_{\nu^+}(q^{-\rho})s_{\nu^3}(q^{-\nu^+-\rho}). 
\end{aligned}
     \label{Schur_expansion_fermion_rep}
\end{eqnarray}
\end{proposition} 

\paragraph{\it Proof\/.}
By inserting the identity 
$\sum_{\nu\in\mathcal{P}}|\nu\rangle\langle\nu|=1$ 
between all pairs of adjacent operators in 
$\Gamma_+(\bm{x})\Gamma'_+(\bm{y}) 
\mathbb{G}_\emptyset (1)\Gamma_-(\bm{z})$, 
the LHS of (\ref{Schur_expansion_fermion_rep}) reads   
\begin{align}
   & \langle 0|
           \Gamma_+(\bm{x})\Gamma'_+(\bm{y})
           \mathbb{G}_0\Gamma_-(\bm{z})
       |0\rangle 
   \nonumber \\[1mm] 
   & =\langle 0|
           \Gamma_+(\bm{x})
                 \left( 
                       \sum_{\nu^1\in\mathcal{P}}
                       |\nu^1\rangle\langle\nu^1|
                 \right)
          \Gamma'_+(\bm{y})
                 \left( 
                      \sum_{\nu^+\in\mathcal{P}}
                      |\ltrans{\,\nu}^+\rangle\langle\ltrans{\,\nu}^+|
                 \right)
          \mathbb{G}_\emptyset (1) 
                \left(
                     \sum_{\nu^3\in\mathcal{P}}
                     |\ltrans{\,\nu}^3\rangle\langle\ltrans{\,\nu}^3|
                \right)
          \Gamma_-(\bm{z})
       |0\rangle 
   \nonumber \\[1mm]
   & =\sum_{\nu^1,\nu^+,\nu^3\in\mathcal{P}}
             s_{\nu^1}(\bm{x}) 
             s_{\nu^+/\ltrans{\,\nu}^1}(\bm{y})
             s_{\ltrans{\,\nu}^3}(\bm{z})  
          \langle\ltrans{\,\nu}^+|
                \mathbb{G}_\emptyset(1)
          |\ltrans{\,\nu}^3\rangle.       
    \label{Schur_expansion_fermion_rep_proof1}
\end{align}
Note that (\ref{matrix_element_Gamma}) and (\ref{matrix_element_Gamma'}) 
have been used in the last line.

The matrix element of 
$\mathbb{G}_\emptyset(1) = 
q^{-K/2}\Gamma_-(q^\rho)\Gamma_+(q^\rho)q^{K/4}$ 
can be expressed in terms of special values of the Schur function as    
\begin{align}
       \langle\ltrans{\,\nu}^+| 
          \mathbb{G}_\emptyset(1)
       |\ltrans{\,\nu}^3\rangle
    =q^{\kappa(\nu^+)+\kappa(\nu^3 )/4}
           s_{\nu^+}(q^{-\rho})s_{\nu^3}(q^{-\nu^+-\rho}). 
    \label{C_two_leggs}
\end{align}
Pluging (\ref{C_two_leggs}) into 
the RHS of (\ref{Schur_expansion_fermion_rep_proof1}), 
we obtain (\ref{Schur_expansion_fermion_rep}). 

Let us derive (\ref{C_two_leggs}). 
It is convenient to use the formula \cite{Takasaki_Nakatsu_JPA_2016}
\begin{align}
    q^{K/2}\Gamma_-(q^{-\rho})\Gamma_+(q^{-\rho}) 
          |\ltrans{\,\alpha}\rangle  
               = \Gamma'_- (q^{-\alpha-\rho}) 
                    |0\rangle s_{\,\ltrans{\,\alpha}}(q^{-\rho}), \quad  
    \forall\,\alpha\in\mathcal{P}. 
  \label{exchange_formula}
\end{align} 
By taking the transpose and using (\ref{exchange_formula}),  
we can rewrite the LHS of (\ref{C_two_leggs}) as  
\begin{align*}
  \langle\ltrans{\,\nu}^+|\mathbb{G}_\emptyset(1)|\ltrans{\,\nu}^3\rangle
     & =q^{\kappa(\nu^+)/2}
           \langle\ltrans{\,\nu}^+| 
              \Gamma_-(q^{-\rho})\Gamma_+(q^{-\rho})q^{K/2}
           |\ltrans{\,\nu}^3\rangle q^{\kappa(\nu^3)/4} 
     \nonumber \\[2mm]
     & =q^{\kappa(\nu^+)/2} 
           \left( 
               s_{\ltrans{\,\nu}^+}(q^{-\rho})
                 \langle 0|\Gamma'_+(q^{-\nu^+-\rho})
           \right)
                |\ltrans{\,\nu}^3\rangle 
           q^{\kappa(\nu^3)/4} 
     \nonumber \\[2mm]
     & =q^{\kappa(\nu^+)+\kappa(\nu^3)/4}
          s_{\nu^+}(q^{-\rho})s_{\nu^3}(q^{-\nu^+-\rho}).  
\end{align*} 
Note that the identity 
$s_{\ltrans{\,\nu}^+}(q^{-\rho})=q^{\kappa(\nu^+)/2}s_{\nu^+}(q^{-\rho})$ 
is used in the last line. Thus, we obtain (\ref{C_two_leggs}).
\qed \\

\begin{remark} 
$\Gamma_-(q^{-\rho})\Gamma_+(q^{-\rho})$ is a self-adjoint operator, 
that is, satisfies      
\begin{align}
    \langle\ltrans{\,\beta}|
          \Gamma_-(q^{-\rho})\Gamma_+(q^{-\rho})
    |\ltrans{\,\alpha}\rangle
        =\langle\ltrans{\,\alpha}|
             \Gamma_-(q^{-\rho})\Gamma_+(q^{-\rho})
          |\ltrans{\,\beta}\rangle, 
                \quad\forall\,\alpha,\beta\in\mathcal{P}. 
    \label{two-legged_identity_matrix}
\end{align}
The matrix elements in (\ref{two-legged_identity_matrix}) can be expressed 
in terms of special values of the Schur function by (\ref{exchange_formula}). 
Actually, the LHS of (\ref{two-legged_identity_matrix}) becomes 
\begin{align*} 
    & \langle\ltrans{\,\beta}|
              \Gamma_-(q^{-\rho})\Gamma_+(q^{-\rho})
        |\ltrans{\,\alpha}\rangle 
          =q^{\kappa(\beta)/2}
            \langle\ltrans{\,\beta}|
                q^{K/2}\Gamma_-(q^{-\rho})\Gamma_+(q^{-\rho})
            |\ltrans{\,\alpha}\rangle  
    \\[2mm] 
    & \qquad \qquad 
    = q^{\kappa(\beta)/2} 
       \langle\ltrans{\,\beta}|
           q^{K/2} \Gamma'_-(q^{-\alpha-\rho})
       |0\rangle 
       s_{\,\ltrans{\,\alpha}}(q^{-\rho})
    = q^{\frac{\kappa(\alpha)+\kappa(\beta)}{2}}  
       s_\alpha(q^{-\rho})s_\beta(q^{-\alpha-\rho}), 
\end{align*}
whereas the RHS of (\ref{two-legged_identity_matrix}) reads  
\begin{align*}
      \langle\ltrans{\,\alpha}|
           \Gamma_-(q^{-\rho})\Gamma_+(q^{-\rho})
      |\ltrans{\,\beta}\rangle 
        =q^{\frac{\kappa(\alpha)+\kappa(\beta)}{2}} 
             s_\alpha(q^{-\beta-\rho})s_\beta(q^{-\rho}). 
\end{align*}
Thus, (\ref{two-legged_identity_matrix}) implies the identity 
\begin{align}
       s_\alpha(q^{-\rho})s_\beta(q^{-\alpha-\rho}) 
            =s_\alpha(q^{-\beta-\rho}) s_\beta (q^{-\rho}),  
       \quad \forall\,\alpha,\beta\in\mathcal{P}.
    \label{two-legged_identity_Schur}
\end{align}
The identity (\ref{two-legged_identity_Schur}) is exploited in 
\cite{Takasaki_Nakatsu_JPA_2016} 
to simplify open topological string amplitudes 
including the case of open topological string on closed vertex.  \\
\end{remark} 

\begin{remark} 
From (\ref{fermion_expression_topological_vertex}), 
the matrix element of $\mathbb{G}_\emptyset(1)$ 
can be written in terms of the two-legged topological vetrex as 
\begin{align*}
    \langle\ltrans{\,\nu}^+|
          \mathbb{G}_\emptyset(1)
    |\ltrans{\,\nu}^3\rangle
    =q^{\kappa(\nu^+)-\kappa(\nu^3)/4} 
            \mathcal{C}_{\left(\ltrans{\,\nu}^+,\emptyset,\,\nu^3\right)}(q). 
\end{align*}
On the other hand, (\ref{topological_vertex_def}) and (\ref{C_two_leggs}) 
yield another expression
\begin{align*}
   \langle\ltrans{\,\nu}^+|
       \mathbb{G}_\emptyset(1)
   |\ltrans{\,\nu}^3\rangle 
    = q^{\kappa(\nu^+)-\kappa(\nu^3)/4} 
        \mathcal{C}_{\left(\nu^3,\ltrans{\,\nu}^+,\emptyset\right)}(q). 
\end{align*}
Furthermore, 
by (\ref{topological_vertex_def}) and (\ref{two-legged_identity_Schur}), 
we find 
\begin{align*}
    & \mathcal{C}_{\left(\nu^3,\ltrans{\,\nu}^+,\emptyset\right)}(q) 
            =q^{\kappa(\nu^3)/2} s_{\nu^3}(q^{-\nu^+-\rho}) s_{\nu^+} (q^{-\rho}) 
  \nonumber \\[1.5mm] 
    & \qquad \qquad
             =q^{\kappa(\nu^3)/2} s_{\nu^3}(q^{-\rho}) s_{\nu^+} (q^{-\nu^3-\rho})  
             =\mathcal{C}_{\left(\emptyset, \nu^3,\ltrans{\,\nu}^+\right)}(q).
\end{align*}
These identities imply the cyclic symmetry of the two-legged topological vertex: 
\begin{align*}
   \mathcal{C}_{\left(\ltrans{\,\nu}^+,\emptyset,\,\nu^3\right)}(q) 
       =\mathcal{C}_{\left(\nu^3,\ltrans{\,\nu}^+,\emptyset\right)}(q) 
       =\mathcal{C}_{\left(\emptyset,\nu^3,\ltrans{\,\nu}^+\right)}(q). 
\end{align*}
\end{remark} 

\subsection{Proof of Theorem \ref{theorem_conjecture}} 

We note that the exponentiation of quadratic form of the power sums 
in (\ref{W(q;x,y,z;1)_fermion_expression}) has a Schur function expansion 
of the following form: 
\begin{proposition} 
\begin{align}
    \exp\left( 
              \sum_{m=1}^\infty\frac{(-1)^{m+1}}{m}p_m(\bm{x})p_{2m}(\bm{z})
           \right) 
       =\sum_{\eta^1,\eta^3\in\mathcal{P}} 
               s_{\,\ltrans{\,\eta}^1}(\bm{x})s_{\eta^3}(\bm{z}) 
                    \sum_{\xi\in\mathcal{P}}
                        \frac{\chi_{\eta^1}(\xi)\chi_{\eta^3}(2\xi)}{z_\xi}.  
    \label{eq3_proof_conjecture}
\end{align}
\end{proposition} 

\paragraph{\it Proof\/.}
We rewrite the LHS of  (\ref{eq3_proof_conjecture}) into  
\begin{align} 
    & \exp\left( 
                  \sum_{m=1}^\infty\frac{(-1)^{m+1}}{m}p_m(\bm{x})p_{2m}(\bm{z})
               \right) 
  \nonumber \\[1.5mm]
    & \quad
        =\prod_{k=1}^\infty 
          \exp\left( 
                 \frac{(-1)^{k-1}}{k}p_k(\bm{x})p_{2k}(\bm{z})
                 \right)  
        =\prod_{k=1}^\infty 
          \sum_{m_k=0}^\infty
                \frac{(-1)^{(k-1)m_k}}{m_k! k^{m_k}} 
                      \left(p_k(\bm{x})p_{2k}(\bm{z})\right)^{m_k} 
  \nonumber \\[1.5mm]
     & \quad   
        =\sum_{m_1,m_2,\cdots}
                  \frac{(-1)^{\sum_{k=1}^\infty (k-1)m_k}}{\prod_{k=1}^\infty m_k!k^{m_k}} 
                         \prod_{k=1}^\infty\bigl(p_k(\bm{x})p_{2k}(\bm{z})\bigr)^{m_k} 
        =\sum_{\xi\in\mathcal{P}} 
                   \frac{(-1)^{|\xi|-l(\xi)}}{z_\xi}p_\xi(\bm{x})p_{2\xi}(\bm{z}), 
     \label{eq4_proof_conjecture}
\end{align}
where the summation over non-negative integers $m_k$'s  
is replaced with the summation over partitions $\xi$  
by arranging $\xi=(1^{m_1}2^{m_2}\cdots)$. 
Plugging  the expansions 
\begin{align*}
  (-1)^{|\xi|-l(\xi)}p_\xi(\bm{x}) 
       =\sum_{\eta^1\in\mathcal{P}} 
            \chi_{\eta^1}(\xi)s_{\,\ltrans{\,\eta}^1}(\bm{x}), \quad  
  p_{2\xi}(\bm{z}) 
       =\sum_{\eta^3\in\mathcal{P}}
             \chi_{\eta^3}(2\xi)s_{\eta^3}(\bm{z})
\end{align*}
into the RHS of (\ref{eq4_proof_conjecture}), 
we obtain (\ref{eq3_proof_conjecture}). 
\qed 

\bigskip

We can now rewrite $\mathcal{W}\left(q;\bm{x},\bm{y},\bm{z};1\right)$ as follows. 
\begin{proposition} 
\label{prop:proof_conjecture} 
\begin{align} 
   \mathcal{W}\left(q;\bm{x},\bm{y},\bm{z};1\right) 
     & =\sum_{\overrightarrow{\mu}\in\mathcal{P}^3}
                     \mathcal{\widetilde{C}}_{\overrightarrow{\mu}}(\lambda)
                         q^{-\kappa(\mu^{(1)})/2+\kappa(\mu^{(2)})+\kappa(\mu^{(3)})/4} 
                 s_{\mu^{(1)}}(\bm{x})s_{\mu^{(2)}}(\bm{y})s_{\mu^{(3)}}(\bm{z}), 
   \label{eq1_proof_conjecture} 
\end{align}
where $\mathcal{\widetilde{C}}_{\overrightarrow{\mu}}(\lambda)$ 
is given by (\ref{LLLZ_formula}) with $q=e^{-\sqrt{-1}\lambda}$.  \\
\end{proposition} 

Theorem \ref{theorem_conjecture} is 
an immediate consequence of Proposition \ref{prop:proof_conjecture}. 
By (\ref{generating_function_topological_vertex}),  
$\mathcal{W}\left(q;\bm{x},\bm{y},\bm{z};1\right)$ is a generating function 
of $\mathcal{C}_{\overrightarrow{\mu}}(q)$ of the form 
\begin{align*}
  \mathcal{W}\left(q;\bm{x},\bm{y},\bm{z};1\right) 
       =\sum_{\overrightarrow{\mu}\in\mathcal{P}^3}
               \mathcal{C}_{\overrightarrow{\mu}}(q) 
                     q^{-\kappa(\mu^{(1)})+\kappa(\mu^{(2)})/2-\kappa(\mu^{(3)})/4} 
                s_{\mu^{(1)}}(\bm{x})s_{\mu^{(2)}}(\bm{y})s_{\mu^{(3)}}(\bm{z}).
\end{align*}
Equating the coefficients of the Schur function products in this expression 
with those of (\ref{eq1_proof_conjecture}), we obtain (\ref{tildeC=C}). 
This also implies that (\ref{eq1_theorem_conjecture}) holds. 

\paragraph{\it Proof of Proposition \ref{prop:proof_conjecture}\/.}  
We derive an expansion of the fermionic representation 
(\ref{W(q;x,y,z;1)_fermion_expression}) in the basis of the Schur functions.  
By (\ref{Schur_expansion_fermion_rep}) and (\ref{eq3_proof_conjecture}),  
$\mathcal{W}\left(q;\bm{x},\bm{y},\bm{z};1\right)$ can be expressed as    
\begin{align} 
   & \mathcal{W}\left(q;\bm{x},\bm{y},\bm{z};1\right) 
   \nonumber \\[1mm]
   & = \sum_{\nu^1,\nu^3,\nu^+\in\mathcal{P}}
         s_{\nu^1}(\bm{x}) 
         s_{\nu^+/\ltrans{\,\nu}^1}(\bm{y})
         s_{\,\ltrans{\,\nu}^3}(\bm{z}) 
             q^{\kappa(\nu^+)+\kappa(\nu^3)/4}
         s_{\nu^+}(q^{-\rho})s_{\nu^3}(q^{-\nu^+-\rho}) 
   \nonumber \\
   & \qquad\qquad \qquad \times 
          \sum_{\eta^1,\eta^3\in\mathcal{P}} 
                     s_{\,\ltrans{\,\eta}^1}(\bm{x}) 
                     s_{\eta^3}(\bm{z}) 
          \sum_{\xi\in\mathcal{P}} 
                     \frac{\chi_{\eta^1}(\xi)\chi_{\eta^3}(2\xi)}{z_\xi} 
   \nonumber \\[1.5mm]
   & = \sum_{\nu^1,\nu^3,\nu^+,\eta^1,\eta^3\in\mathcal{P}} 
          s_{\,\ltrans{\,\eta}^1}(\bm{x}) 
          s_{\nu^1}(\bm{x}) 
          s_{\nu^+/\ltrans{\,\nu}^1}(\bm{y})
          s_{\eta^3}(\bm{z})
          s_{\,\ltrans{\,\nu}^3}(\bm{z}) 
   \nonumber \\
   & \qquad\qquad\qquad \times 
               q^{\kappa(\nu^+)+\kappa(\nu^3)/4} 
               s_{\nu^+}(q^{-\rho})s_{\nu^3}(q^{-\nu^+-\rho})
     \sum_{\xi\in\mathcal{P}}
               \frac{\chi_{\eta^1}(\xi)\chi_{\eta^3}(2\xi)}{z_\xi}. 
 \label{eq5_proof_conjecture}
\end{align}
The products of the Schur funcions and the skew Schur function 
can be expressed in terms of 
the Littlewood-Richardson numbers as 
\begin{align*} 
  s_{\,\ltrans{\,\eta}^1}(\bm{x}) s_{\nu^1}(\bm{x}) 
       & =\sum_{\mu^{(1)}\in\mathcal{P}}
                      c_{\,\ltrans{\,\eta}^1\nu^1}^{~\mu^{(1)}}
                                 s_{\mu^{(1)}}(\bm{x}), 
  \\[0.5mm]
  s_{\nu^+/\ltrans{\,\nu}^1}(\bm{y})
       & =\sum_{\mu^{(2)}\in\mathcal{P}} 
                      c_{\,\ltrans{\,\nu}^1\mu^{(2)}}^{~\nu^+}
                                 s_{\mu^{(2)}}(\bm{y}), 
  \\[0.5mm]
  s_{\eta^3}(\bm{z})s_{\,\ltrans{\,\nu}^3}(\bm{z})
       & =\sum_{\mu^{(3)}\in\mathcal{P}}
                      c_{\eta^3\,\ltrans{\,\nu}^3}^{~\mu^{(3)}}
                                 s_{\mu^{(3)}}(\bm{z}).
\end{align*} 
Consequently, 
\begin{align*}
  & \mathcal{W}\left(q;\bm{x},\bm{y},\bm{z};1\right) 
     \\[0.5mm]
  & =\sum_{\nu^1,\nu^3,\nu^+,\eta^1,\eta^3\in\mathcal{P}} 
        \Biggl(  
            \sum_{\mu^{(1)}\in\mathcal{P}} 
                c_{\,\ltrans{\,\eta}^1\nu^1}^{~\mu^{(1)}} 
                s_{\mu^{(1)}}(\bm{x})
        \Biggr) 
        \Biggl(
            \sum_{\mu^{(2)}\in\mathcal{P}} 
                c_{\,\ltrans{\,\nu}^1\mu^{(2)}}^{~\nu^+}
                s_{\mu^{(2)}}(\bm{y})
        \Biggr) 
        \Biggl(  
            \sum_{\mu^{(3)}\in\mathcal{P}}
                c_{\eta^3\,\ltrans{\,\nu}^3}^{~\mu^{(3)}}
                s_{\mu^{(3)}}(\bm{z})
        \Biggr)  \\
  & \qquad\qquad\qquad \times 
           q^{\kappa(\nu^+)+\kappa(\nu^3)/4}
           s_{\nu^+}(q^{-\rho})s_{\nu^3}(q^{-\nu^+-\rho})
          \sum_{\xi\in\mathcal{P}}
          \frac{\chi_{\eta^1}(\xi)\chi_{\eta^3}(2\xi)}{z_\xi} \\
  & =\sum_{(\mu^{(1)},\mu^{(2)},\mu^{(3)})\,\in\mathcal{P}^3} 
           s_{\mu^{(1)}}(\bm{x})
           s_{\mu^{(2)}}(\bm{y})
           s_{\mu^{(3)}}(\bm{z})  \\
  & \times 
         \sum_{\nu^1,\nu^3,\nu^+,\eta^1,\eta^3\in\mathcal{P}} 
          c_{\,\ltrans{\,\eta}^1\nu^1}^{~\mu^{(1)}}
          c_{\,\ltrans{\,\nu}^1\mu^{(2)}}^{~\nu^+} 
          c_{\eta^3\,\ltrans{\,\nu}^3}^{~\mu^{(3)}} 
          q^{\kappa(\nu^+)+\kappa(\nu^3)/4}
          s_{\nu^+}(q^{-\rho})s_{\nu^3}(q^{-\nu^+-\rho}) 
          \sum_{\xi\in\mathcal{P}}
          \frac{\chi_{\eta^1}(\xi)\chi_{\eta^3}(2\xi)}{z_\xi}.  
\end{align*}
Thus we obtain (\ref{eq1_proof_conjecture}).  
\qed \\

\section{Proof of Theorem \ref{theorem_NKdV}} 

\subsection{Factorization formulas at $\tau=1/N$}

Just like $\mathbb{G}[\bm{t};1]$ and $\mathbb{G}'[\bm{t};1]$, 
the operator-valued generating functions  (\ref{G[t;tau]}) and (\ref{G'[t;tau]}) 
specialized to $\tau=1/N$ ($N=1,2,\dots$) can be factorized 
to a triple product of operators. 
\begin{theorem}{\sf(factorization formulas at $\tau=1/N$)} 
\label{factorization_formulas_1/N} 
Let $N$ be a positive integer. 
The operator-valued generating functions 
$\mathbb{G}[\bm{t};\tau]$ and $\mathbb{G}'[\bm{t};\tau]$ 
satisfy the following identities at $\tau=1/N$: 
\begin{gather} 
     \mathbb{G}\bigl[\bm{t};1/N\bigr]
           =\exp\left( 
                        \sum_{k=1}^\infty t_kq^{kJ_0}J_{kN}
                   \right)
             \mathbb{G}_\emptyset \bigl(1/N\bigr)
             \exp\left( 
                        \sum_{k=1}^\infty 
                             (-1)^{k+1}t_kq^{kJ_0}J_{k(N+1)} 
                   \right),  
        \label{factorization_1/N} 
     \\[3mm]
     \mathbb{G}'\bigl[\bm{t};1/N\bigr]
           =\exp\left( 
                        \sum_{k=1}^\infty 
                              (-1)^{kN+1}t_kq^{-kJ_0}J_{kN} 
                    \right) 
             \mathbb{G}'_\emptyset \bigl(1/N\bigr) 
             \exp\left( 
                        \sum_{k=1}^\infty 
                              (-1)^{kN}t_kq^{-kJ_0}J_{k(N+1)}
                   \right).  
       \label{factorization'_1/N}
\end{gather}
\end{theorem} 

\paragraph{\it Proof\/.}
The proof is mostly parallel to 
the proof of Theorem \ref{factorization_formulas_tau=1}.  
We derive differential equations that 
the operator-valued generating functions satisfy at $\tau=1/N$. 
We describe the case of $\mathbb{G}\left[\bm{t};1/N\right]$ in detail. 
Partial derivatives of $\mathbb{G}\left[\bm{t};1/N\right]$ 
with respect to $\bm{t}$ 
can be expressed in the form of (\ref{partial_t_G[t;tau]}).  
The double sum in (\ref{partial_t_G[t;tau]}) can be computed by 
Proposition \ref{ext_shift_symmetry_A_tau}. 
By specializing $\tau$ to $\tau=1/N$, 
(\ref{ext_shift_symmetry_a_tau}) reads 
\begin{align}
   J_{-kN}\mathbb{G}_\alpha(1/N)
          = \mathbb{G}_\alpha(1/N)(-1)^kJ_{-k(N+1)} 
             +q^{kJ_0}
                \sum_{\beta\in\mathcal{R}_{k,\alpha}} 
                      \mbox{\sf sgn}(\alpha,\beta) 
                      \mathbb{G}_\beta(1/N)  
   \label{ext_shift_symmetry_c1}
\end{align}
for $k\in\mathbb{Z}, k \neq 0$.  
By letting $k=-m$, we have   
\begin{align}
     \sum_{\beta\in\mathcal{R}^{(+)}_{m,\alpha}} 
                 {\sf sgn}(\alpha,\beta) 
                 \mathbb{G}_\beta(1/N) 
        =q^{mJ_0}J_{mN} 
            \mathbb{G}_\alpha(1/N)
        +\mathbb{G}_\alpha(1/N)(-1)^{m+1}q^{m J_0}J_{m(N+1)}
   \label{ext_shift_symmetry_c2}
\end{align}  
for $m=1,2,\dots$\,. 
The RHS of (\ref{partial_t_G[t;tau]}) thereby becomes 
\begin{align*}
  & \mbox{RHS of (\ref{partial_t_G[t;tau]}) at $\displaystyle \tau\!=\!1/N$} 
  \\[1mm]
  & \qquad 
     =\sum_{\alpha\in\mathcal{P}} 
           \Bigl( 
               q^{mJ_0}J_{mN}\mathbb{G}_\alpha (1/N) 
               +\mathbb{G}_\alpha(1/N)(-1)^{m+1}q^{mJ_0}J_{m(N+1)}
           \Bigr) 
       s_\alpha[\bm{t}] 
   \\[1mm]
   & \qquad 
      =q^{mJ_0}J_{mN}
             \mathbb{G}[\bm{t};1/N]
        +\mathbb{G}[\bm{t};1/N](-1)^{m+1}q^{mJ_0}J_{m(N+1)}. 
\end{align*} 
We thus find that $\mathbb{G}\left[\bm{t};1/N\right]$ 
satisfies the first-order differential equations 
\begin{align}
  \partial_{t_m}\mathbb{G}[\bm{t};1/N]
      =q^{mJ_0}J_{mN} 
             \mathbb{G}[\bm{t};1/N]
       +\mathbb{G}[\bm{t};1/N](-1)^{m+1}q^{mJ_0}J_{m(N+1)}
  \label{PDE_G[t;1/N]}
\end{align} 
for $m=1,2,\dots$\,. 
General solutions of the differential equations (\ref{PDE_G[t;1/N]}) 
are given by 
\begin{align}
  \mathbb{G}[\bm{t};1/N] 
    = \exp\left( 
                 \sum_{k=1}^\infty t_kq^{kJ_0}J_{kN} 
              \right) 
      \mathbb{G}[\bm{0};1/N]
      \exp\left( 
                \sum_{k=1}^\infty (-1)^{k+1}t_kq^{kJ_0}J_{k(N+1)} 
             \right). 
  \label{GS_PDE_G[t;1/N]}
\end{align}
Since $\mathbb{G}[\bm{0};1/N]=\mathbb{G}_\emptyset(1/N)$,  
(\ref{GS_PDE_G[t;1/N]}) yields (\ref{factorization_1/N}).  

(\ref{factorization'_1/N}) can be likewise derived 
from Proposition \ref{ext_shift_symmetry_B_tau}.  
Partial derivatives of $\mathbb{G}'[\bm{t};1]$ with respect to $\bm{t}$ 
take the form of (\ref{partial_t_G'[t;tau]}). 
The double sum therein can be computed using 
Proposition \ref{ext_shift_symmetry_B_tau}. 
By letting $\tau=1/N$, (\ref{ext_shift_symmetry_b_tau}) 
yields       
\begin{align}
   \sum_{\beta\in\mathcal{R}^{(+)}_{m,\alpha}}
                {\sf sgn}(\alpha,\beta)\mathbb{G}'_\beta(1/N) 
   = (-1)^{mN}\Bigl( 
                     -q^{-mJ_0}J_{mN}\mathbb{G}'_\alpha(1/N)
                     +\mathbb{G}'_\alpha(1/N)q^{-mJ_0}J_{m(N+1)}
                   \Bigr)
  \label{ext_shift_symmetry_d2}
\end{align} 
for $m=1,2,\dots$\,. 
By plugging (\ref{ext_shift_symmetry_d2}) 
into the RHS of (\ref{partial_t_G'[t;tau]}), 
we eventually find out that 
$\mathbb{G}'[\bm{t};1/N]$ satisfies 
the first-order differential equations   
\begin{align}
  & \partial_{t_m}\mathbb{G}'[\bm{t};1/N] 
  \nonumber \\[1mm]
  & \quad =(-1)^{mN+1}
                q^{-mJ_0}J_{mN}
                \mathbb{G}'[\bm{t};1/N]
                +\mathbb{G}'[\bm{t};1/N] 
                (-1)^{mN}q^{-mJ_0}J_{m(N+1)}
   \label{PDE_G'[t;1/N]}
\end{align}
for $m=1,2,\dots$\,. 
General solutions of (\ref{PDE_G'[t;1/N]}) are given by 
\begin{align}
  & \mathbb{G}'[\bm{t};1/N] 
  \nonumber \\
  & \quad =\exp\left( 
                          \sum_{k=1}^\infty(-1)^{kN+1}t_kq^{-kJ_0}J_{kN}
                       \right)
                \mathbb{G}'[\bm{0};1/N] 
                \exp\left( 
                         \sum_{k=1}^\infty(-1)^{kN}t_kq^{-kJ_0}J_{k(N+1)}
                      \right). 
    \label{GS_PDE_G'[t;1/N]}
\end{align}
Since $\mathbb{G}'[\bm{0};1/N]=\mathbb{G}'_\emptyset(1/N)$, 
(\ref{GS_PDE_G'[t;1/N]}) yields the formula (\ref{factorization'_1/N}). 
\qed 

\subsection{Representation of generating function at $\tau=N$}

The foregoing factorization formula gives a new fermionic representation 
of the generating function of the topological vertex. 
\begin{proposition} 
\label{fermionic_formula_W(q;x,y,z;N)} 
Let $N$ be a positive integer. 
The generating function (\ref{generating_function_topological_vertex}) 
at $\tau=N$ has a fermionic representation of the form   
\begin{align}
  \mathcal{W}\left(q;\bm{x},\bm{y},\bm{z};N\right) 
    & =\langle 0| 
             \exp\left(
                       \sum_{k=1}^\infty \frac{p_k(\bm{x})}{k}J_{kN}
                   \right)
        \Gamma'_+(\bm{y}) 
        \mathbb{G}_\emptyset \bigl(1/N\bigr)\Gamma_-(\bm{z}) 
        |0\rangle 
  \nonumber \\
    & \quad \times 
        \exp\left( 
                 \sum_{m=1}^\infty 
                     \frac{(-1)^{m+1}}{m}
                     p_m(\bm{x})p_{m(N+1)}(\bm{z})\right).
  \label{W(q;x,y,z;N)_fermion_expression}
\end{align}
\end{proposition} 

\paragraph{\it Proof\/.}
The cyclic symmetry (\ref{cyclicity_topological_vertex}) can be embodied  
in the language of the generating functions 
(\ref{generating_function_topological_vertex}) as  
\begin{align*}
    \mathcal{W}\bigl(q;\bm{x},\bm{y},\bm{z};\tau\bigr)
       =\mathcal{W}\Bigl(q;\bm{z},\bm{x},\bm{y};\frac{-1}{1+\tau}\Bigr). 
\end{align*} 
In view of (\ref{W(q;x,y,z;tau)_fermion_expression}), 
this relation leads to the expression 
\begin{align}
      \mathcal{W}(q;\bm{x},\bm{y},\bm{z};\tau)
       =\langle 0|\Gamma_+(\bm{z})
            \left(
                \sum_{\alpha\in\mathcal{P}}
                   \mathbb{G}_\alpha\Bigl(\frac{-1}{1+\tau}\Bigr) 
                   s_{\,\ltrans{\,\alpha}}(\bm{x})
            \right)
        \Gamma'_-(\bm{y})|0\rangle. 
\label{W(q;x,y,z;N)_fermion_representation_1}
\end{align}
We can replace the VEV $\langle 0|\,\dots\,|0\rangle$ in the RHS of 
(\ref{W(q;x,y,z;N)_fermion_representation_1}) with the transpose. 
Actually, since the operator-valued functions 
$\mathbb{G}_\alpha (-1/1+\tau)$ satisfy
\begin{align*}
    \langle\beta|
            \mathbb{G}_\alpha\Bigl(\frac{-1}{1+\tau}\Bigr)
    |\gamma\rangle 
    =\langle\gamma| 
            \mathbb{G}_{\,\ltrans{\,\alpha}}\bigl(1/\tau\bigr)
     |\beta\rangle, 
   \quad \forall \,\beta,\gamma\in\mathcal{P}, 
\end{align*}   
we can rewrite (\ref{W(q;x,y,z;N)_fermion_representation_1}) as   
\begin{align*}
    \mathcal{W}(q;\bm{x},\bm{y},\bm{z};\tau)
         =\langle 0|\Gamma'_+(\bm{y})
               \left( 
                      \sum_{\alpha\in\mathcal{P}} 
                      \mathbb{G}_\alpha\bigl(1/\tau\bigr) 
                      s_{\alpha}(\bm{x})
               \right)
           \Gamma_-(\bm{z})|0\rangle.  
\end{align*}
In particular, by specialization to $\tau=N$, we find  
\begin{align}
   \mathcal{W}(q;\bm{x},\bm{y},\bm{z};N)
        =\langle 0|\Gamma'_+(\bm{y})
                \left(
                       \sum_{\alpha\in\mathcal{P}} 
                       \mathbb{G}_\alpha \bigl(1/N\bigr) 
                       s_{\alpha}(\bm{x})
                \right)
          \Gamma_-(\bm{z})|0\rangle.  
   \label{W(q;x,y,z;N)_fermion_representation_2}
\end{align}

By substituting $t_k=p_k(\bm{x})/k$, 
(\ref{factorization_1/N}) converts the sum of operators 
in (\ref{W(q;x,y,z;N)_fermion_representation_2}) 
into a triple product of operators as    
\begin{align*} 
   & \mathcal{W}(q;\bm{x},\bm{y},\bm{z};N) 
\\[1mm]
   & =\langle 0|\Gamma'_+(\bm{y})
              \exp\left(
                            \sum_{k=1}^\infty\frac{p_k(\bm{x})}{k}J_{kN
                    }\right)
              \mathbb{G}_\emptyset\bigl(1/N\bigr) 
              \exp\left(
                            \sum_{k=1}^\infty
                            \frac{(-1)^{k+1}p_k(\bm{x})}{k}J_{k(N+1)}
                     \right)
         \Gamma_- (\bm{z})|0\rangle 
\\[1mm]
   & =\langle 0|
              \exp\left(
                          \sum_{k=1}^\infty \frac{p_k(\bm{x})}{k}J_{kN}
                     \right)
              \Gamma'_+(\bm{y}) 
              \mathbb{G}_\emptyset \bigl(1/N\bigr)
              \Gamma_-(\bm{z})
        |0\rangle 
        \exp\left(
                      \sum_{m=1}^\infty 
                         \frac{(-1)^{m+1}}{m}
                           p_m(\bm{x})p_{m(N+1)}(\bm{z})
               \right). 
\end{align*}
We thus obtain (\ref{W(q;x,y,z;N)_fermion_expression}). 
\qed

\subsection{Proof of Theorem \ref{theorem_NKdV}}
We first derive a reduction formula of the generating functions 
of three-partition Hodge integrals at positive integral values of $\tau$. 
This formula gives a generalization of (\ref{reduction}) 
at all positive integral values of $\tau$. 
We then consider integrable hierarchies that capture a certain aspect of 
integrable structure underlying the generating functions 
of two-partition Hodge integrals. 
By combining these considerations, 
we eventually obtain Theorem \ref{theorem_NKdV}. 

\bigskip

Let $N$ be a positive integer. 
\begin{proposition} 
\label{reduction_formula_G_tau=N} 
The generating function (\ref{connected_G}) satisfies the relation
\begin{align}
  G\bigl(\lambda;(p^{(1)},p^{(2)},p^{(3)});N\bigr) 
      =G\bigl(\lambda;(0,p^+,p^{(3)});N\bigr)
         + \sum_{m=1}^\infty\frac{(-1)^{m+1}}{m}p^{(1)}_mp^{(3)}_{m(N+1)}, 
   \label{reduction_G_tau=N}
\end{align}
where $p^+=(p^+_k)_{k=1}^\infty$ is a linear combination of 
$p^{(1)}$ and $p^{(2)}$ given by 
\begin{align}
    p^+_k 
    & =\left\{\begin{array}{cl} 
        (-1)^{k+1}Np^{(1)}_{k/N}+p^{(2)}_k, & \mbox{$k\equiv 0$ mod $N$},
        \\[2mm]
        p^{(2)}_k, &\mbox{$k\not\equiv 0$ mod $N$}. 
         \end{array}\right.
            \label{p+_tau=N}
\end{align}
\end{proposition}  

\paragraph{\it Proof\/.}
By Theorem \ref{theorem_conjecture} and 
Proposition \ref{fermionic_formula_W(q;x,y,z;N)}, 
we can rewrite the exponentiated generating function as   
\begin{align}
    & \exp\left(G(\lambda;\overrightarrow{p};N)\right) 
    \nonumber \\[2mm]
    & \quad = \langle 0|\exp\left(\sum_{k=1}^\infty \frac{p^{(1)}_k}{k}J_{kN}
          + \sum_{k=1}^\infty\frac{(-1)^{k+1}p^{(2)}_k}{k}J_{k}\right)
                \mathbb{G}_\emptyset \bigl(1/N\bigr)
            \exp\left(\sum_{k=1}^\infty\frac{p^{(3)}_k}{k}J_{-k}\right)|0\rangle 
    \nonumber \\[0.5mm]
    & \qquad\qquad  \times 
       \exp\left(\sum_{m=1}^\infty\frac{(-1)^{m+1}}{m}p^{(1)}_mp^{(3)}_{m(N+1)}\right) 
    \nonumber \\[2mm]
    & \quad = \langle 0|\exp\left(\sum_{k=1}^\infty\frac{(-1)^{k+1}p^+_k}{k}J_{k}\right) 
                \mathbb{G}_\emptyset\bigl(1/N\bigr)
            \exp\left(\sum_{k=1}^\infty \frac{p^{(3)}_k}{k}J_{-k}\right)|0\rangle 
     \nonumber \\[0.5mm]
    & \qquad\qquad \times 
       \exp\left(\sum_{m=1}^\infty\frac{(-1)^{m+1}}{m}p^{(1)}_mp^{(3)}_{m(N+1)}\right) .
             \label{exp(G)_tau=N}
\end{align}
By letting $\overrightarrow{p}=(0,p^{(2)},p^{(3)})$, 
(\ref{exp(G)_tau=N}) takes the simplified form   
\begin{align*}
   & \exp\left(G(\lambda;(0,p^{(2)},p^{(3)});N)\right) \\[1.0mm]
   &\qquad = \langle 0|\exp\left(\sum_{k=1}^\infty\frac{(-1)^{k+1}p^{(2)}_k}{k}J_{k}\right)
               \mathbb{G}_\emptyset\bigl(1/N\bigr)
                           \exp\left(\sum_{k=1}^\infty \frac{p^{(3)}_k}{k}J_{-k}\right)|0\rangle. 
\end{align*}
We can replace the VEV 
$\langle 0|\,\dots\,|0\rangle$ in the last part of (\ref{exp(G)_tau=N}) 
with this expression. This yields the relation 
\begin{align*}
     & \exp\left(G(\lambda;(p^{(1)},p^{(2)},p^{(3)});N)\right) \\ 
     & \qquad = \exp\left(G(\lambda;(0,p^+,p^{(3)});N)\right) 
           \exp\left(\sum_{m=1}^\infty\frac{(-1)^{m+1}}{m}p^{(1)}_mp^{(3)}_{m(N+1)}\right),
\end{align*}
hence (\ref{reduction_G_tau=N}). 
\qed \\

\begin{proposition} 
\label{integrable_structure_two_partition_tau=N} 
$\calT(\lambda,N,0,p^{(2)},\bst)$ is a tau function 
of the KP hierarchy with respect to the time variables $\bst$, 
and satisfies the equations 
\begin{align}
   \frac{\partial^2\log\calT (\lambda,N,0,p^{(2)},\bst)}{\partial t_k\partial t_{m(N+1)}}
    = 0,\quad k,m = 1,2,\ldots. 
\label{531}
\end{align}
\end{proposition} 

\paragraph{\it Proof\/.} 
We start from the fermionic expression 
\begin{equation}
  \calT(\lambda,N,0,p^{(2)},\bst) 
  = \langle 0|\mathbf{g}(\lambda,N;p^{(2)})
    \exp\left(\sum_{k=1}^\infty t_kJ_{-k}\right)|0\rangle,
  \label{531proof-1}
\end{equation}
where $\mathbf{g}(\lambda,N;p^{(2)})$ is an element of $GL(\infty)$ of the form 
\begin{equation*}
  \mathbf{g}(\lambda,N;p^{(2)}) 
  = \exp\left(\sum_{k=1}^\infty\frac{(-1)^{k+1}p^{(2)}_k}{k}J_k\right)
    \mathbb{G}_\emptyset(1/N). 
\end{equation*}
(\ref{ext_shift_symmetry_c1}) contains the special subset 
\begin{equation*}
  \mathbb{G}_\emptyset(1/N)J_{-m(N+1)} 
  = (-1)^mJ_{-mN}\mathbb{G}_\emptyset(1/N), \quad m = 1,2,\ldots, 
\end{equation*}
which yields the algebraic relation 
\begin{gather*}
  \mathbf{g}(\lambda,N;p^{(2)})\exp\left(\sum_{m=1}^\infty t_{m(N+1)}J_{-m(N+1)}\right)  
  \notag\\[0.5mm]
  = \exp\left(\sum_{m=1}^\infty(-1)^mt_{m(N+1)}J_{-mN}\right)
    \mathbf{g}(\lambda,N;p^{(2)})
    \exp\left(- \sum_{m=1}^\infty(-1)^{m(N+1)}p^{(2)}_{mN}t_{m(N+1)}\right) 
\end{gather*}
satisfied by $\mathbf{g}(\lambda,N;p^{(2)})$.  
We can use this algebraic relation to rewrite (\ref{531proof-1}) as 
\begin{align*}
  \calT(\lambda,N,0,p^{(2)},\bst) 
  & = \langle 0|\mathbf{g}(\lambda,N;p^{(2)})
    \exp\left(\sum_{r=1}^N\sum_{m=1}^\infty t_{m(N+1)-r}J_{-m(N+1)+r}
    \right)|0\rangle 
  \notag\\
  & \quad\mbox{}\times
    \exp\left(- \sum_{m=1}^\infty(-1)^{m(N+1)}p^{(2)}_{mN}t_{m(N+1)}\right). 
\end{align*}
This implies that $\calT(\lambda,N,0,p^{(2)},\bst)$ satisfies (\ref{531}). 
\qed \\

\paragraph{\it Proof of Theorem \ref{theorem_NKdV}\/.}
By (\ref{reduction_G_tau=N}), 
$\calT(\lambda,N,p^{(1)},p^{(2)},\bst)$ can be factorized as 
\begin{equation*}
\begin{aligned}
  \calT(\lambda,N,p^{(1)},p^{(2)},\bst) 
  &= \exp\left(G(\lambda;\,0,p^{+},p^{(3)};\,N)\right)
   \exp\left(\sum_{m=1}^\infty\frac{(-1)^{m+1}}{m}p^{(1)}_mp^{(3)}_{m(N+1)}\right)\\
  &= \calT(\lambda,N,0,p^{+},\bst)
   \exp\left(\sum_{m=1}^\infty\frac{(-1)^{m+1}}{m}p^{(1)}_mp^{(3)}_{m(N+1)}\right). 
\end{aligned}
\end{equation*}
This relation and (\ref{531}) imply that 
$\calT(\lambda,N,p^{(1)},p^{(2)},\bst)$ satisfies (\ref{red-cond2}). 
\qed \\



\begin{thebibliography}{99}

\bibitem{Graber_Pandharipande_1999}
  T.~Graber and R.~Pandharipande, 
  Localization of virtual classes, 
  Invent.~Math. {\bf 135} (1999) 487--518. 

\bibitem{Kontsevich_1995}
  M.~Kontsevich, 
  Enumeration of rational curves via torus actions, 
  Progr.~Math. {\bf 129}, Birkh\"{a}user, Boston, Boston, MA, 1995, 
  pp. 335--368. 

\bibitem{Li_Liu_Liu_Zhou_2009}
  J.~Li, C.-C.~M.~Liu, K.~Liu and J.~Zhou, 
  A mathematical theory of the topological vertex,    
  Geom.~Topol. {\bf 13} (2009), 527--621,
  arXiv:math/0408426v3. 

\bibitem{AKMV}
  M.~Aganagic, A.~Klemm, M.~Mari$\tilde{\mbox{n}}$o and C.~Vafa, 
  The topological vertex, 
  Commun.~Math.~Phys. {\bf 254} (2005), 425--478, 
  arXiv:hep-th/0305132.  

\bibitem{Liu_Liu_Zhou_2004}
  C.-C.~Liu, K.~Liu and J.~Zhou, 
  On a proof of a conjecture of Mari\~{n}o-Vafa on Hodge Integrals,
  Math.~Res.~Lett. {\bf 11} (2004), no. 2-3, 259--272, 
  arXiv:math/0306257.

\bibitem{Liu_Liu_Zhou_2003}
  C.-C.~Liu, K.~Liu and J.~Zhou, 
  A proof of a conjecture of Mari\~{n}o--Vafa on Hodge Integrals, 
  J.~Differential Geom. {\bf 65} (2003), 289--340, 
  arXiv:math/0306434. 

\bibitem{Liu_Liu_Zhou_2007}
  C.-C.~M.~Liu, K.~Liu and J.~Zhou, 
  A formula of two-partition Hodge integrals, 
  J.~Amer.~Math.~Soc.~{\bf 20} (2007), no. 1, 149--184, 
  arXiv:math/0310272v3.

\bibitem{Marino_Vafa_2001}
  M.~Mari\~{n}o and C.~Vafa, 
  Framed knots at large N, 
  Contemp.~Math. {\bf 310} 
  (Amer. Math. Soc., Providence, RI, 2002), pp. 185--204,  
  arXiv:hep-th/0108064. 

\bibitem{ELSV_2001}
  T.~Ekedahl, S.~Lando, M.~Shapiro and A.~Vainshtein, 
  Hurwitz numbers and intersections on moduli spaces of curves, 
  Invent.~Math. 1{\bf 46} (2001), 297--327, 
  arXiv:math/0004096.

\bibitem{Nakatsu_Takasaki_CMP_2009} 
  T.~Nakatsu and K.~Takasaki, 
  Melting crystal, quantum torus and Toda hierarchy, 
  Comm.~Math.~Phys. {\bf 285} (2009), 445--156,  
  arXiv:0710.5339 [hep-th].

\bibitem{Nakatsu_Takasaki_ASPS} 
  T.~Nakatsu and K.~Takasaki, 
  Integrable structure of melting crystal model with external potentials, 
  Adv.~Stud.~Pure~Math. vol.~{\bf 59}, 201--223  
  (Mathematical Society of Japan, Tokyo, 2010),  
  arXiv:0807.4970 [math-ph].




\bibitem{Takasaki_Nakatsu_JPA_2016}
  K.~Takasaki and T.~Nakatsu, 
  Open string amplitudes of closed topological vertex, 
  J.~Phys. A: ~Math.~Theor. {\bf 49} (2016), 025403,  
  arXiv:1507.07053 [math-ph]. 

\bibitem{Takasaki_2014}
  K.~Takasaki, 
  Generalized Ablowitz-Ladik hierarchy in topological string theory, 
  J.~Phys. A: ~Math.~Theor. {\bf 47} (2014), 165201,  
  arXiv:1312.7184 [math-ph]. 

\bibitem{Takasaki_Nakatsu_SIGMA_2017}
  K.~Takasaki and T.~Nakatsu, 
  $q$-difference Kac-Schwarz operators in topological string theory, 
  SIGMA {\bf 13} (2017), 009,  
  arXiv:1609.00882 [math-ph].

\bibitem{Nakatsu_Takasaki_2019}
  T.~Nakatsu and K.~Takasaki, 
  to appear.

\bibitem{ADKMV}
  M.~Aganagic, R.~Dijkgraaf, A.~Klemn, M.~Mari$\tilde{\mbox{n}}$o and C.~Vafa, 
  Topological strings and integrable hierarchies,  
  Commun.~Math.~Phys. {\bf 261} (2006), 451--516, 
  arXiv:hep-th/0312085.  

\bibitem{Zhou_2010}
  J.~Zhou,  Hodge integrals and integrable hierarchies, 
  Lett.~Math.~Phys. {\bf 93} (2010), 55--71, 
  arXiv:math/0310408. 

\bibitem{Macdonald_book}
   I.~G Macdonald, Symmetric functions and Hall polynomials, 
  Oxford University Press 1995. 

\bibitem{Iqbal-Kzcaz-Vafa} 
   A.~Iqbal and C.~Kz$\mbox{\c{c}}$az and C.~Vafa, 
   The refined topological vertex, 
  JHEP {\bf 0910} (2009), 069, 
   arXiv:hep-th/0701156. 

\bibitem{Taki}
   M.~Taki, Refined topological vertex and instanton counting, 
   JHEP {\bf 0803} (2008), 048,  
   arXiv:0710.1776 [hep-th]. 

\bibitem{Dickey_book}
   L.~A.~Dickey, Soliton Equations and Hamiltonian System, 
   World Scientific, Singapore, 2003. 

\bibitem{Goulden_Jackson_1997}
   I.~P.~Goulden and D.~M.~Jackson, 
   Transitive factorizations into transpositions and 
   holomorphic mappings on the sphere, 
   Proc.~Amer.~Math.~Soc. {\bf 125} (1997), 51-60. 

 




\end{thebibliography}
\end{document}